\documentclass[twocolumn]{aastex631}
\usepackage{amsmath}

\begin{document}

\title{Evolution of Extremely Soft Binaries in Dense Star Clusters: On the Jupiter Mass Binary Objects}

\author{Yihan Wang}
\affiliation{Nevada Center for Astrophysics, University of Nevada, 4505 S. Maryland Pkwy., Las Vegas, NV 89154-4002, USA}
\affiliation{Department of Physics and Astronomy, University of Nevada, 4505 S. Maryland Pkwy., Las Vegas, NV 89154-4002, USA}

\author{Rosalba Perna}
\affiliation{Department of Physics and Astronomy, Stony Brook University, Stony Brook, NY 11794-3800, USA}

\author{Zhaohuan Zhu}
\affiliation{Nevada Center for Astrophysics, University of Nevada, 4505 S. Maryland Pkwy., Las Vegas, NV 89154-4002, USA}
\affiliation{Department of Physics and Astronomy, University of Nevada, 4505 S. Maryland Pkwy., Las Vegas, NV 89154-4002, USA}

\author{Douglas N. C. Lin}
\affiliation{Department of Astronomy \& Astrophysics, University of California, Santa Cruz, CA 95064, USA}
\affiliation{Institute for Advanced Studies, Tsinghua University, Beijing 100086, China}

%%\collaboration{20}{(AAS Journals Data Editors)}

%% Note that the \and command from previous versions of AASTeX is now
%% depreciated in this version as it is no longer necessary. AASTeX 
%% automatically takes care of all commas and "and"s between authors names.

%% AASTeX 6.31 has the new \collaboration and \nocollaboration commands to
%% provide the collaboration status of a group of authors. These commands 
%% can be used either before or after the list of corresponding authors. The
%% argument for \collaboration is the collaboration identifier. Authors are
%% encouraged to surround collaboration identifiers with ()s. The 
%% \nocollaboration command takes no argument and exists to indicate that
%% the nearby authors are not part of surrounding collaborations.

%% Mark off the abstract in the ``abstract'' environment. 
\begin{abstract}
Star-forming regions, characterized by dense environments, experience frequent encounters that significantly influence binary systems, leading to their hardening, softening, or ionization. We extend the Hut \& Bahcall formalism to derive an analytical expression for the ionization cross-section in extreme mass ratio binary systems, allowing us to investigate the orbital evolution and population dynamics of binary planets and binary brown dwarfs in star clusters, while considering ongoing binary system formation.
Our findings reveal that for low-mass soft binaries, the semi-major axis distribution asymptotes to a universal power law between $\propto a^{-8/3}$ and $\propto a^{-5/3}$ over the derived ionization timescale.
We also discuss the implications of our results for the candidate Jupiter-mass binary objects putatively reported in the Trapezium cluster.
We demonstrate that if their existence is verified, they likely form continuously with a spectrum proportional to $a^{1}$, aligning better with the ejection mechanism than with the in-situ formation mechanism, which predicts a distribution roughly proportional to $a^{-1}$. However, this implies an impractically high ejection formation rate. Alternatively, if these objects are binary brown dwarfs, continuous in-situ formation ($\propto a^{-1}$) with an initial minimal semi-major axis around 20 AU and a formation rate of 100 Myr$^{-1}$ plausibly matches the observed number of single objects, binary number, binary fraction, and semi-major axis distribution.
\end{abstract}

%% Keywords should appear after the \end{abstract} command. 
%% The AAS Journals now uses Unified Astronomy Thesaurus concepts:
%% https://astrothesaurus.org
%% You will be asked to selected these concepts during the submission process
%% but this old "keyword" functionality is maintained in case authors want
%% to include these concepts in their preprints.

%% From the front matter, we move on to the body of the paper.
%% Sections are demarcated by \section and \subsection, respectively.
%% Observe the use of the LaTeX \label
%% command after the \subsection to give a symbolic KEY to the
%% subsection for cross-referencing in a \ref command.
%% You can use LaTeX's \ref and \label commands to keep track of
%% cross-references to sections, equations, tables, and figures.
%% That way, if you change the order of any elements, LaTeX will
%% automatically renumber them.
%%
%% We recommend that authors also use the natbib \citep
%% and \citet commands to identify citations.  The citations are
%% tied to the reference list via symbolic KEYs. The KEY corresponds
%% to the KEY in the \bibitem in the reference list below. 

\section{Introduction} \label{sec:intro}
Binary stars in clusters play a crucial role in cluster evolution. Without binaries, clusters would collapse into infinite density through a process known as core collapse, driven by two-body relaxation. However, interactions between single stars and hard binaries in star clusters transfer binding energy from the binaries to the surrounding single stars via binary hardening, which ‘heats up’ the cluster and alleviates or even stops core collapse.  Although the interactions between binaries and single stars are complicated by the three-body problem, significant advancements have been made analytically based on impulsive or secular approximations \citep{Heggie1975, Hut1983, hut83b, hut1984, hut1993, Heggie1993}. 
  
Numerous studies have explored the steady-state distribution of binary binding energy in star clusters. In an ideal thermal equilibrium state, the properties of these binaries can be directly derived from classical statistical mechanics, independent of the binary interaction rate \citep{Ambartsumian1937,Lynden-Bell1969, Heggie1975}. In such an equilibrium, the rate of binary creation would balance the rate of binary ionization, and the rate of binary hardening would balance the rate of softening. However, achieving this equilibrium is unlikely as the equilibrium distribution function of binary binding energy diverges at both the hard and soft ends. While several works (e.g. \citealt{Lightman1978,Retterer1980}) have examined the non-equilibrium distribution at the high binding energy (hard) end, soft binaries are often overlooked in binary distribution studies because they are quickly ionized in star clusters.

Extending beyond binary stars, another line of research, especially active following the discovery of exoplanets \citep{Mayor1995}, 
has focused on the evolution of planetary systems harbored in dense star clusters
(e.g. \citealt{Heggie1996,spurzem2009,
Portegies2015,Li2015,Malmberg2011,Shara2016,Cai2017,Cai2018,Cai2019,Rice2018,Li2019,Li2020,Wang2020a,Wang2020b}).
Planet-star binaries, with their large mass ratios, differ significantly from binary stars. The cross-sections of planetary system-star interactions have been extensively discussed by  \citet{Hills1989,Hills1990}
\citet{Sigurdsson1992,Sigurdsson1993}. 

The dynamics of planetary systems in star clusters has recently garnered increased interest due to the growing number of free-floating planets discovered in nearby star formation regions and clusters \citep{Scholz2012,Miret-Roig2022}, as well as to a variety of unusual planetary architectures that cannot be naturally explained by standard planet formation and migration theories\citep{Lin1996, Zhu2012, Zhu2019, Chen2020, Li2024}. It is believed that some of these free-floating planets originated from planet-planet scattering \citep{Gladman1993, zhoulin2007,
Juric2008, Chatterjee2008, idalin2013} 
or planetary systems within star clusters and were ejected by stellar flybys (e.g. \citealt{Malmberg2011,  Wang2015}). Similarly, unusual planetary systems could be the outcome of  significant perturbation by flyby stars. In particular, the discovery of hot Jupiters in star clusters has demonstrated that stellar flybys can induce high-eccentricity migration in planetary systems, leading to the formation of hot Jupiters (e.g. \citealt{Shara2016, wang2020c, Rodet2021,wang2022}). All of these discoveries make the evolution of planet-star binaries in star clusters particularly intriguing.

More interestingly, recent observations with the James Webb Space Telescope (JWST) have revealed a significant population of Jupiter-mass objects (JMO) in the Trapezium cluster of the Orion Nebula. Among these ~540 JMOs, about 40 pairs are observed moving together \citep{Pearson2023}. These Jupiter-Mass Binary Objects (JuMBOs) have attracted considerable attention because their formation is not easily explained by current theoretical models. Some efforts to explain the formation of JuMBOs involve pair ejection from planetary systems \citep{Wang2024, Lazzoni2024, Yu2024}
or in-situ formation \citep{Portegies2023}. Each mechanism has its own formation efficiency and distribution of orbital parameters for the planetary binaries. Once formed, these systems undergo further dynamical interactions, resulting in softening, hardening, and ionization, which together determine the time-evolution of the JuMBO properties.

Given their very small binding energy compared to the kinetic energy of surrounding stars, these JuMBOs are extremely soft in the Trapezium cluster. The evolution and distribution function of such extremely soft binaries in star clusters could be significantly different from those of binary stars or planet-star binaries. The dynamics of these low-mass binaries in star clusters is therefore of considerable interest, and it is the subject of this investigation.

Our first step (\S2) into computing this evolution is a generalization of the ionization cross section in the extreme mass ratio limit. Our analytically derived cross-section, validated with numerical experiments, is accurate for extreme mass ratios. We then use (\S3) the derived cross section to explore the long-term evolution of the orbital separation of the binaries which survive, uncovering an asymptotic behaviour of such distribution. Last (\S4 \& \S5), we discuss how our results can be used to set stringent limits on observable JuMBO/PMO populations if one imposes coupled constraints on their absolute number,  binary fraction, and semi-major axis distribution. We summarize and conclude in \S6. 

\section{Scattering between an extremely soft binary and a single star}

Binaries in star clusters can be divided into two categories: the soft binary regime, where the binding energy of the binaries, consisting of objects of masses $m_1$ and $m_2$, is smaller than the average kinetic energy of the surrounding stars (of mass $m_3$), and the hard binary regime, where the binding energy of the binary is larger than the average kinetic energy of the surrounding stars.

For equal-mass cases in the soft binary regime, where the stars in binaries and single stars have equal masses, the orbital velocity of the binaries is much smaller than the velocity of the single flyby stars during the scattering process. Therefore, scatterings between soft binaries and single stars are usually addressed using the so-called impulsive approximation. In this approximation, during the scattering process, the positions of the stars in binaries are assumed to be unchanged, and only velocity changes are estimated.

Depending on the impact parameter $b$ between the center of mass of the binary and the flyby star, the scatterings can be further divided into two regimes. If the impact parameter $b$ is much smaller than the semi-major axis of the binary $a_{12}$, meaning only one component of the binary closely interacts with the flyby star, the velocity change of the further component can be ignored. If the impact parameter $b$ is much larger than the semi-major axis of the binary $a_{12}$, both velocity kicks of the two components in the binary need to be considered. For equal mass cases
(the mass of each binary component $m_1=m_2$ equals to that of the flyby star
$m_3$) the binding energy of the binaries can only be significantly changed if $b$ is much smaller than $a_{12}$. Thus, in scenarios involving significant binding energy changes, binary component swap, and binary ionization, the scattering process is usually dealt with in the close flyby regime where $b < a_{12}$ \citep{Hut1983}.

However, if the masses of both components in the binary are much smaller than the mass of the flyby star ($m_1 \simeq m_2 \ll m_3$), {the binaries become even softer compared to the equal mass case.} In this regime, the flyby stars do not necessarily need to be close to the binary to significantly alter its binding energy. Therefore, in this extremely soft regime where the impulsive approximation becomes more accurate, both velocity kicks of $m_1$ and $m_2$ become relevant and need to be included in order to study binary evolution.

\begin{figure}
    \includegraphics[width=\columnwidth]{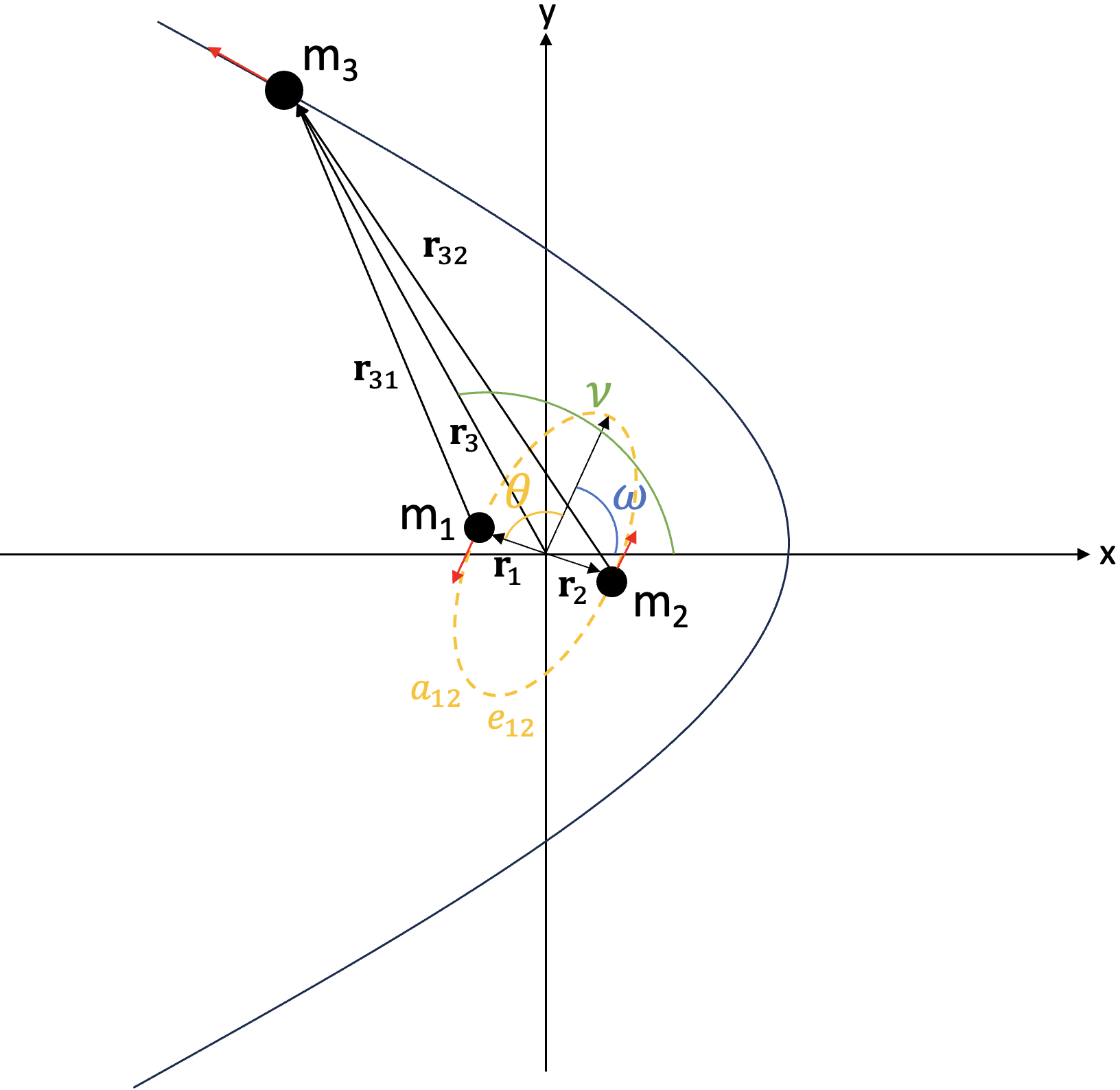}
    \caption{Schematics of single-binary scatterings: Massive object $m_3$ encounters  a binary consisting of the small objects $m_1$ and $m_2$ with semi-major axis $a_{12}$ and eccentricity $e_{12}$.  The instantaneous positions of $m_1$, $m_2$ and $m_3$ are $\mathbf{r}_1$, $\mathbf{r}_2$, and $\mathbf{r}_3$, respectively. }
    \label{fig:schematics}
\end{figure}

To calculate the scattering between an extremely soft binary and a single star using the impulsive approximation, we establish our coordinate system at the center of mass of the soft binary as shown in Fig.~\ref{fig:schematics}. The velocities and positions of the three objects can be expressed as follows:

\begin{eqnarray}
r_1&=&\frac{m_2}{m_{12}}\frac{a_{12}(1-e_{12}^2)}{1+e_{12}\cos\theta}\\
&=&\frac{m_2}{m_{12}}\frac{l_{12}}{1+e_{12}\cos\theta}=\frac{l_1}{1+e_{12}\cos\theta}\\
r_2&=&\frac{m_1}{m_{2}}r_1\\
r_3&=&\frac{l_3}{1+e_3\cos\nu}\\
\mathbf{r}_{31} &=& (r_3\cos\nu- r_1\cos\bar{\omega}, r_3\sin\nu- r_1\sin\bar{\omega})\\
\mathbf{r}_{32} &=& (r_3\cos\nu+ r_2\cos\bar{\omega}, r_2\sin\nu+ r_1\sin\bar{\omega})\\
\mathbf{v}_1&=&\frac{m_2}{m_{12}}\sqrt{\frac{Gm_{12}}{l_{12}}}\nonumber\\
&&(-\sin\bar\omega-e_{12}\sin\omega,\cos\bar\omega+e_{12}\cos\omega)\\
\mathbf{v}_2&=&-\frac{m_1}{m_{2}}\mathbf{v}_1\,,
\end{eqnarray}
where $e_{12}$, $\theta$, and $l_{12}$ are the eccentricity, true anomaly, and semi-latus rectum of the inner Keplerian orbit ($m_1$, $m_2$), respectively; $e_3$ ($>$1), $\nu$, and $l_3 = \frac{h_3^2}{Gm_{123}}$ are the eccentricity, true anomaly, and semi-latus rectum of the outer Keplerian orbit ([$m_1$, $m_2$], $m_3$), respectively. $\bar{\omega} = \theta + \omega$ is the argument of the latitude, $m_{12}=m_1+m_2$ $m_{123}=m_1+m_2+m_3$, and $h_3 = bv_\infty$ is the specific angular momentum of the hyperbolic orbit.

The velocity kick on $m_1$ and $m_2$ can then be estimated via
\begin{eqnarray}
\delta\mathbf{v}_i &=& \frac{1}{m_i}\int_{-\infty}^{+\infty}\mathbf{F}_{3i}(t)dt\\
&=&\frac{1}{m_i}\int_{-\nu_\infty}^{+\nu_\infty}\mathbf{F}_{3i}(\nu)\frac{dt}{d\nu}d\nu\,,
\end{eqnarray}
where
\begin{eqnarray}
\mathbf{F}_{3i}&=&\frac{-Gm_3m_i}{r_{3i}^3}\mathbf{r}_{3i}\\
\nu_\infty &=& \arccos(1/e_3)\\
\frac{dt}{d\nu} &=& \frac{r_3^2}{\sqrt{Gm_{123}l_3}} = \frac{r_3^2}{h_3}\,,
\end{eqnarray}
with $\pm\nu_\infty$ being the true anomaly of $m_3$ at $t=\pm\infty$. In the limit of $m_3 \gg m_1$, $m_2$ and $r_3 \gg a_{12}$, and given that $\int_{-\nu_\infty}^{+\nu_\infty} f(\cos\nu)d\cos\nu=0$, we can write\footnote{Details are provided in the Appendix.}

\begin{eqnarray}
&&\delta\mathbf{v}_1 =\frac{Gm_3}{bv_\infty}\times\nonumber\\
&&\left(
\begin{aligned}
&&\frac{2\sqrt{e_3^2-1}}{e_3}+\frac{r_1}{l_3}\cos\bar{\omega}\left((2+\frac{5}{e_3^2})\sqrt{e_3^2-1}+\nu_\infty\right)\\
&&\frac{r_1}{l_3}\sin\bar{\omega}\left(\nu_\infty-\frac{5}{e^2_3}\sqrt{e_3^2-1}\right)
\end{aligned}
\right)\label{eq:v1}\\
&&\delta\mathbf{v}_2 =\frac{Gm_3}{bv_\infty}\times\nonumber\\
&&\left(
\begin{aligned}
&&\frac{2\sqrt{e_3^2-1}}{e_3}-\frac{r_2}{l_3}\cos\bar{\omega}\left((2+\frac{5}{e_3^2})\sqrt{e_3^2-1}+\nu_\infty\right)\\
&&-\frac{r_2}{l_3}\sin\bar{\omega}\left(\nu_\infty-\frac{5}{e^2_3}\sqrt{e_3^2-1}\right)\label{eq:v2}
\end{aligned}
\right)
\end{eqnarray}

In the impulsive regime where the positions $m_1$ and $m_2$ remain approximately unchanged during the scattering, we can derive the relative semi-major axis or relative binding energy change $\Delta$, 
\begin{eqnarray}
    \Delta = \frac{\delta\epsilon_{12}}{\epsilon_{12}}=\frac{\delta a_{12}}{a_{12}}&\sim&\frac{a_{12}}{Gm_{12}}(\delta v_{12}^2+2\mathbf{v}_{12}\cdot \delta\mathbf{v}_{12})\nonumber\\
    &\sim& \frac{a_{12}}{Gm_{12}}2\mathbf{v}_{12}\cdot \delta\mathbf{v}_{12}\,,
    \label{eq:delta}
\end{eqnarray}
having used
\begin{eqnarray}
    -\frac{Gm_{12}}{2a_{12}}=\frac{1}{2}v_{12}^2 -\frac{Gm_{12}}{r_{12}}\,,
\end{eqnarray}
where 
\begin{eqnarray}
\mathbf{v}_{12}&=&\mathbf{v}_{2}-\mathbf{v}_{1}\\
\delta\mathbf{v}_{12}&=&\delta\mathbf{v}_{2}-\delta\mathbf{v}_{1}
\end{eqnarray}
and $\epsilon_{12}= \frac{Gm_{1}m_2}{2a_{12}}$ is the binding energy of binary $m_{12}$. Plugging Eq.~\ref{eq:v1} and \ref{eq:v2} into Eq.~\ref{eq:delta}, we obtain
\begin{eqnarray}
\Delta &\sim&\frac{m_3}{m_{12}}\frac{m_{123}}{m_{12}}\left(\frac{a_{12}}{b}\right)^3\left(\frac{v_{\rm orb}}{v_\infty}\right)^3\Phi(\theta,\omega,e_{12},e_3)\,,
\end{eqnarray}
where $v_{\rm orb} = \sqrt{\frac{Gm_{12}}{a_{12}}}$ is the circular orbital velocity of the binary $m_{12}$, and 
\begin{eqnarray}
\Phi(\theta,\omega,e)&=& \sqrt{1-e_{12}^2}\left(2+\frac{10}{e_3^2}\right)\sqrt{e_{3}^2-1}\sin2\bar\omega\nonumber\\
&+&2e_{12}\sqrt{1-e_{12}^2} \bigg(\nu_\infty\sin\theta - \frac{5}{e_3^2}\sqrt{e_3^2-1}\sin(\omega+\bar\omega)\nonumber\\
&+& 2\sqrt{e_3^2-1}\sin\omega\cos\bar\omega\bigg)
\end{eqnarray}
is the phase term.

Then the differential cross-section $\frac{d\sigma}{d\Delta}$ of the scattering that changes the relative binding energy of the binary by $\Delta$ in the leading order of $\Delta$ can be expressed as follows:

\begin{eqnarray}
    \frac{d\sigma}{d\Delta} &=& \sum_{b>0} 2\pi b\left|\frac{db}{d\Delta}\right|\nonumber\\
    &\sim&\frac{2\pi}{3}\left(\frac{Gm_{12}}{v_\infty}\right)\left(\frac{m_{123}}{m_{12}}\right)^{4/3}\frac{a_{12}}{|\Delta|^{5/3}}\Phi^{2/3}\nonumber\\
    &\sim&\frac{2\pi}{3}\left(\frac{Gm_{12}}{v_\infty^2}\right)\left(\frac{m_{3}}{m_{12}}\right)^{4/3}\frac{a_{12}}{|\Delta|^{5/3}}\,.\label{eq:diff-cross-section}
\end{eqnarray}
This relation applies to both binary softening, where $\Delta<0$, and binary hardening, where $\Delta>0$. One may note that this expression becomess divergent as $\Delta\rightarrow0$. Indeed, in star clusters, there is a maximum physical impact parameter, i.e. the inter-particle distance $R_{\rm int}\sim \left(\frac{1}{n}\right)^{1/3}$, where $n$ is the stellar number density of the cluster. This leads to 
\begin{eqnarray}
    |\Delta|_{\rm min} = \frac{m_3}{m_{12}}\frac{m_{123}}{m_{12}}na^3\left(\frac{v_{\rm orb}}{v_\infty}\right)^3\,.\label{eq:delta-min}
\end{eqnarray}
For ionizations where $\Delta <-1$, the total ionization cross section is
\begin{eqnarray}
\sigma_{\rm ion} &\sim& \int_{-\infty}^{-1} \frac{d\sigma}{d\Delta} d\Delta\nonumber\\
&=&\pi\left(\frac{Gm_{12}}{v_\infty^2}\right)\left(\frac{m_{3}}{m_{12}}\right)^{4/3}a_{12}\,.
\end{eqnarray}
This is different from the equal mass case derived by \cite{Hut1983},
\begin{eqnarray}
    \sigma_{\rm ion}\sim \pi\left(\frac{Gm_{12}}{v_\infty^2}\right)\left(\frac{m_{3}}{m_{12}}\right)^{2}a_{12}\,.\label{eq:hut}
\end{eqnarray}
Indeed, the ionization cross section derived here is intuitive. Since the binary is in the extremely soft regime, it can be ionized by the tidal force from the flyby star as the pericenter $r_p$ approaches the tidal disruption radius $r_t\sim\left(\frac{m_3}{m_{12}}\right)^{1/3}a_{12}$. The corresponding impact parameter $b_c$ that leads to the binary ionization can then be obtained from the relation
\begin{eqnarray}
b^2 = r_p^2+\frac{2Gm_{123}}{v_\infty^2}r_p\,,
\end{eqnarray}
from which the ionization cross section can be roughly estimated as
\begin{eqnarray}
    \sigma_{\rm ion}&\sim& \pi b_c^2\nonumber\\
    &=&\pi \left(r_t^2+\frac{2Gm_{123}}{v_\infty^2}r_t\right)\nonumber\\
    &=&\pi\left(\left(\frac{m_{123}}{m_{12}}\right)^{2/3}a_{12}^2 + \frac{2Gm_{123}}{v_\infty^2} \left(\frac{m_{123}}{m_{12}}\right)^{1/3}a_{12}\right)\nonumber\\
    &\sim& \pi\left(\frac{Gm_{12}}{v_\infty^2}\right)\left(\frac{m_{3}}{m_{12}}\right)^{4/3}a_{12}\,.
\end{eqnarray}
The last step holds true if gravitational focusing is important, i.e. $r_p^2\ll\frac{2Gm_{123}}{v_\infty^2}r_p$.

In order to verify the expression for  $\sigma_{\rm ion}$ obtained here, we performed few-body scattering experiments with the extremely high precision few-body code {\tt SpaceHub} \citep{Wang2021}. For each mass ratio $q=m_{12}/m_3$ and $v_\infty$, we performed $N=10^6$ scattering experiments with $b^2$ uniformly distributed between [0, $b^2_{\rm max}$], where $b_{\rm max}$ is the maximum impact parameter that ensures that its corresponding closest approach $r_{\rm p, max}$ is $10r_t$. We estimate the binary ionization cross section numerically with the Monte Carlo method
\begin{eqnarray}
    \sigma_{\rm ion} = \pi b^2_{\rm max}\frac{N_{\rm ion}}{N}\,,
\end{eqnarray}
where $N_{\rm ion}$ is the total number of scatterings resulting in binary ionizations.

\begin{figure}
    \includegraphics[width=\columnwidth]{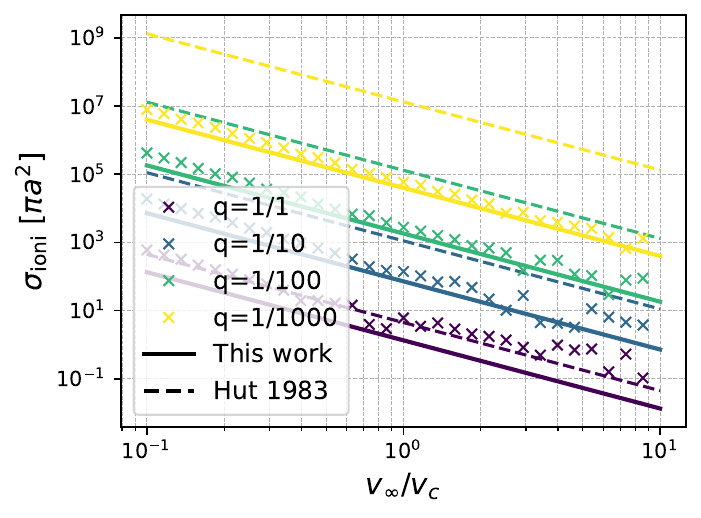}
    \caption{The crosses report the ionization cross section from numerical scattering experiments for different values of the mass ratio $q$ between the binary and the scatterer. Numerical results are contrasted to the analytical formula by \citet{Hut1983} (dashed lines), and the one derived in this work (solid lines). Unlike the former, the latter provides a good match to the numerical results in the low-$q$ regime.}
    \label{fig:verify}
\end{figure}

Fig.~\ref{fig:verify} shows the results of our numerical experiments. It can be seen that, in the equal mass scattering regime, $\sigma_{\rm ion}$ obtained in 
\cite{Hut1983} provides a good approximation, while  $\sigma_{\rm ion}$ derived in this paper underestimates the ionization cross section. However, as the mass ratio $q$ drops, $\sigma_{\rm ion}$ obtained in \cite{Hut1983} becomes inaccurate by orders of magnitude, {even for q=1/10}, while $\sigma_{\rm ion}$ derived in this paper can accurately estimate the ionization cross section. Therefore, to study the long-term evolution of low-mass binaries such as those composed of planets and brown dwarfs, it is of paramount importance to use the cross section derived here.

\section{Evolution of extremely soft binaries in stellar clusters}

Binaries in stellar clusters can be formed via in-situ cloud instabilities, three-body scatterings, two-body scatterings with energy dissipation (e.g., from tidal forces or gravitational wave radiation), and other mechanisms (see e.g. \citealt{PPVII} for a recent review). Each formation mechanism results in binaries with specific distributions in semi-major axis and eccentricity. If the star cluster is not dense enough that the timescale of close flybys becomes shorter than the age of the cluster, binaries will maintain their properties from the time of their formation, assuming the internal evolution of the binaries can be ignored. However, in dense star clusters, frequent stellar flybys will significantly alter the orbital properties of binaries.

For an equal-mass stellar population with stellar mass $m_*$, stellar number density $n$, and velocity dispersion $\sigma_v$, the evolution of the binary distribution $f(\epsilon)$ follows:

\begin{eqnarray}
    \frac{\partial f(\epsilon, t)}{\partial t} &=& \mathcal{C}(\epsilon) - \mathcal{D}(\epsilon)f(\epsilon, t) + \int_0^\infty f(\epsilon^\prime,t)\mathcal{T}(\epsilon^\prime, \epsilon)d\epsilon^\prime \nonumber \\
    &-& f(\epsilon,t)\int_0^\infty \mathcal{T}(\epsilon, \epsilon^\prime)d\epsilon^\prime\,,\label{eq:f}
\end{eqnarray}
where $\mathcal{C}(\epsilon)$ is the creation rate of binaries per unit volume per unit binding energy $\epsilon$, and $\mathcal{C}(\epsilon)$ is assumed to be independent of $f(\epsilon)$ as typical binary formation mechanisms do not require binaries to begin with. $\mathcal{D}(\epsilon)$ is the binary destruction rate per binary, and $\mathcal{T}(\epsilon, \epsilon^\prime)$ is the rate at which a binary with binding energy $\epsilon$ transitions to energies between $\epsilon^\prime$ and $\epsilon^\prime + d\epsilon^\prime$.

The equilibrium distribution can be reached on the timescale of $\sim 1/\mathcal{D}(\epsilon)$ by assuming that the binary creation/destruction and softening/hardening rates are in equilibrium:

\begin{eqnarray}
    \mathcal{C}(\epsilon) &=& \mathcal{D}(\epsilon)f_{\rm eq}(\epsilon)\label{eq:eq} \\
    f_{\rm eq}(\epsilon)\mathcal{T}(\epsilon,\epsilon^\prime) &=& f_{\rm eq}(\epsilon^\prime)\mathcal{T}(\epsilon^\prime, \epsilon)\,,
\end{eqnarray}
{based on the microscopic reversibility\citep{Goodman1993}. }
If we assume that the distribution function of single stars follows a Maxwellian distribution with stellar number density $n$ and velocity dispersion $\sigma_v$, and that particles interact with each other solely through Newtonian gravity (thus, binaries can only be formed via three-body encounters), then statistical mechanics predicts that the equilibrium distribution of the binding energy of binaries formed from a uniform single star population with mass $m_*$ should be \citep{Heggie1975,Goodman1993}:
\begin{eqnarray}
    f_{\rm eq}(\epsilon) \propto \frac{n^2}{\sigma_v^3}\exp\left(\frac{\epsilon}{\kappa}\right)\epsilon^{-5/2}\,,
\end{eqnarray}
where $\kappa = m_*\sigma_v^2$. The corresponding creation (via three-body encounters) and destruction rates follow the scalings \citep{Goodman1993}:

\begin{eqnarray}
    \mathcal{C(\epsilon)} &\sim& \frac{2\pi^2}{3\sqrt{3}}\frac{n^3G^5m_*^4}{\sigma_v^{11}}\frac{(\epsilon/\kappa)^{-9/2}}{(1+\frac{1}{5}\kappa/\epsilon)(1+e^{-\epsilon/\kappa})} \label{eq:create} \\
    \mathcal{D}(\epsilon) &\sim& \frac{8}{3}\sqrt{\frac{\pi}{3}}\frac{nG^2m_*^2}{\sigma_v^3}\frac{(\epsilon/\kappa)^{-2}}{(1+\frac{1}{5}\kappa/\epsilon)(1+e^{\epsilon/\kappa})}\,. \label{eq:des}
\end{eqnarray}
This equilibrium distribution indicates that binaries with binding energy in the equilibrium state follow the power-law distribution $\propto \epsilon^{-5/2}$, or, in terms of semi-major axis, $\propto a^{1/2}$.

It should be noted that this equilibrium distribution diverges in the extremely soft regime as $\epsilon \rightarrow 0$, and cannot be applied to extremely soft binaries for which $m_1, m_2 \ll m_3$. Small-mass soft binaries with the same $\epsilon$ as in the equal mass case do not follow the same mass scaling for the creation rate as described in Eq.~\ref{eq:create}. Indeed, the creation rate via three-body encounters ($\propto m^{4}$, where $m$ is the mass of the binary components\footnote{It is easy to prove that in the unequal mass regime for the same ratio $\epsilon/\kappa$, we can just replace $m_*$ with $m$. The same replacement applies to the destruction rate as well.}) drops rapidly as the mass of the binary components decreases. Thus, small-mass soft binaries are usually formed from other mechanisms following different scalings on $\epsilon$, $n$, $m$, and $\sigma_v$. On the other hand, the destruction rate via binary ionization during close flybys ($\mathcal{D} \propto m^2$) drops much slower. Due to this reason, the condition $\mathcal{C} = \mathcal{D}$ is typically hard to satisfy for soft binaries in the small-mass regime. Therefore, these extremely soft binaries can hardly enter the equilibrium state. In most cases, the distribution function of these extremely soft binaries is in the strong destruction regime where $\mathcal{D} \gg \mathcal{C}$.

\subsection{Evolution of extremely soft binaries}

In this subsection, we focus on solving the distribution function of soft binaries in the small-mass regime as described in Eq.~\ref{eq:f}. Due to the unknown formation mechanism of smal- mass binaries (e.g., Jupiter mass binaries) in stellar clusters, we adopt a parameterized expression for their formation rate as follows,
\begin{eqnarray}
    \mathcal{C}(\epsilon, t) \propto \epsilon^{-\beta} t^{-\alpha}\,.
\end{eqnarray}
This rate can be translated into the more straightforward form
\begin{eqnarray}
    \mathcal{C}(a_{12}, t) \propto a_{12}^{-p} t^{-\alpha}\,,
\end{eqnarray}
where $p = -\beta + 2$, as $\epsilon$ and $a_{12}$ are interchangeable for given $m_1$ and $m_2$. For $\alpha > 1$, binaries are mostly formed at early times; for $-1 < \alpha < 1$, binaries are formed roughly uniformly over time; and for $\alpha < -1$, most binaries are formed at late times.

To simplify the equation, we explore two extreme cases where $\alpha = \infty$ (\textit{diffusion} model), where all binaries form at once at the beginning, and $\alpha = 0$ (\textit{injection} model) where binaries are formed uniformly over time. To avoid divergence, we assume that binaries are formed with binding energy in the range $[\epsilon_{\rm min}, \epsilon_{\rm max}]$ ($[a_{\rm min}, a_{\rm max}]$){, where $\epsilon_{\rm min}/\epsilon_{\rm max}=10^{-3}$}, with the normalization:

\begin{eqnarray}
    &&V\int_0^\infty dt^\prime \int_{\epsilon_{\rm min}}^{\epsilon_{\rm max}} d\epsilon \mathcal{C}(\epsilon,t) = N_0,\quad \alpha=\infty \\
    &&V\int_0^t dt^\prime \int_{\epsilon_{\rm min}}^{\epsilon_{\rm max}} d\epsilon \mathcal{C}(\epsilon,t) = N(t) = \mathcal{R}t,\quad \alpha=0\label{eq:g rate}
\end{eqnarray}
where $V$ is the total volume of the star cluster, $N_0$ is the total number of binaries formed in the \textit{diffusion} model, and $\mathcal{R}$ is the binary formation rate per unit time in the \textit{injection} model. For $p > 0$, most of the binaries are created at $a_{\rm min}$, while for $p < 0$, most of the binaries are created at $a_{\rm max}$. Here we also define $a_{c}$ as the typical size of the created binaries, where

\begin{eqnarray}
    a_c = \left\{ 
    \begin{aligned}
        &&a_{\rm min}, \quad p > 0 \\
        &&a_{\rm max}, \quad p \leq 0\,.
    \end{aligned}
    \right.
\end{eqnarray}
{Therefore, when $p>0$, binaries are injected from $a_c$ to $1000a_c$, when $p<0$, binaries are injected from $a_c/1000$ to $a_c$.}
The transformation rate $\mathcal{T}(\epsilon, \epsilon^\prime)$ in the small mass region can be derived from Eq.~\ref{eq:diff-cross-section}:

\begin{eqnarray}
    \mathcal{T}(\epsilon, \epsilon^\prime) &=& n\sigma_v \left|\frac{d\sigma}{d \epsilon^\prime}(\epsilon, \epsilon^\prime)\right| \nonumber \\
    &=& n\sigma_v \left|\frac{d\sigma}{d\Delta}(\epsilon, \epsilon^\prime)\frac{d\Delta}{d\epsilon^\prime}\right| \nonumber \\
    &=& n\sigma_v \frac{2\pi}{3} \left(\frac{Gm_{12}}{\sigma_v^2}\right) \left(\frac{m_3}{m_{12}}\right)^{4/3} \frac{a_{12}}{\epsilon\left|\frac{\epsilon^\prime - \epsilon}{\epsilon}\right|^{5/3}} \nonumber \\
    &=& \frac{\pi}{12} \frac{nG^2m_{12}^2}{\sigma_v^3} \left(\frac{m_3}{m_{12}}\right)^{1/3} \frac{\kappa}{\epsilon} \frac{\epsilon^{2/3}}{|\epsilon^\prime - \epsilon|^{5/3}}
\end{eqnarray}
where $\kappa = m_3\sigma_v^2$. $\mathcal{T}$ diverges as $\epsilon^\prime \rightarrow \epsilon$. However, due to Eq.~\ref{eq:delta-min}, $\mathcal{T}$ is cut off at $|\Delta|_{\rm min}$. The ionization rate is 
\begin{eqnarray}
    \mathcal{D}(\epsilon) &=& n\sigma_v\sigma_{\rm ion} = \int_{-\infty}^{0}\mathcal{T}(\epsilon, \epsilon^\prime)d\epsilon^\prime \nonumber \\
    &=& \frac{\pi}{8} \frac{nG^2m_{12}^2}{\sigma_v^3} \left(\frac{m_3}{m_{12}}\right)^{1/3} \frac{\kappa}{\epsilon}\,.
\end{eqnarray}
The corresponding ionization timescale is 
\begin{eqnarray}
    \tau(\epsilon) &=& \mathcal{D}^{-1} \nonumber \\
    &=& \frac{8}{\pi} \frac{\sigma_v^3}{nG^2m_{12}^2} \left(\frac{m_{12}}{m_3}\right)^{1/3} \frac{\epsilon}{\kappa} \nonumber \\
    &=& \frac{1}{\pi} \frac{\sigma_v}{nGm_{12}a_{12}} \left(\frac{m_{12}}{m_3}\right)^{4/3}\,.
    \label{eq:tau}
\end{eqnarray}

\begin{figure*}
    \includegraphics[width=2\columnwidth]{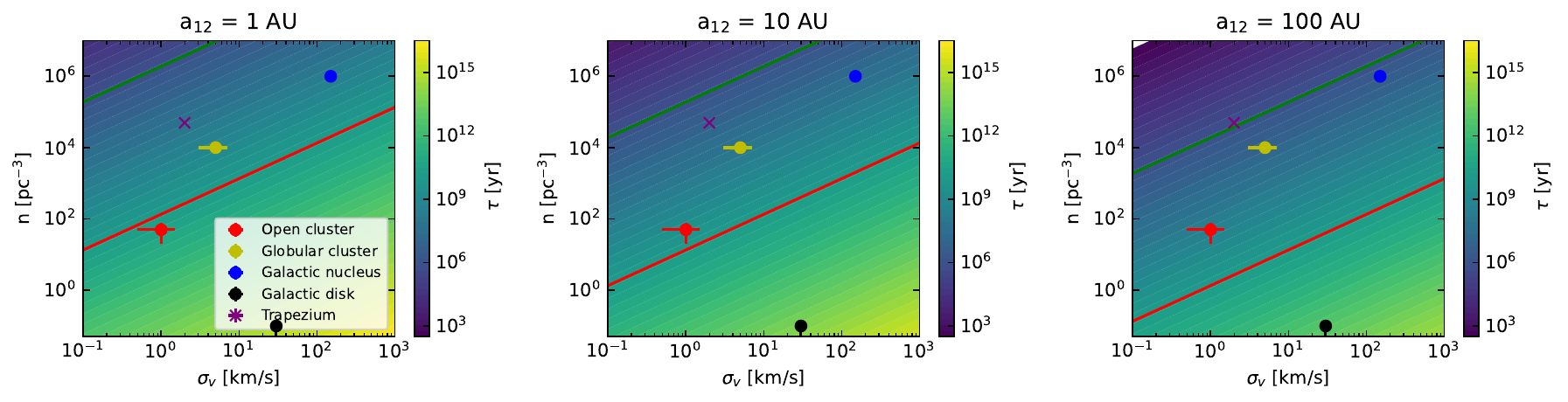}
    \caption{Examples of the ionization time $\tau$ for different values of $n$ and $\sigma_v$ with Jupiter-mass components $m_1 = m_2 = 10^{-3} M_\odot$ and a primary mass $m_3 = 1 M_\odot$. The green solid lines indicate 1 Myr, while the red solid lines represent the Hubble time ($\sim14$~Gyr). The left panel demonstrates that Jupiter-mass binaries with a semi-major axis of 1~AU remain stable for a duration up to the Hubble time in typical open clusters and the Galactic disk. However, their stability is limited to 1~Gyr in typical globular clusters, the Galactic nucleus, and the Trapezium. As shown in the right panel, when the binary size increases to 100~AU, the ionization time for these binaries approximates 1~Myr in globular clusters, the Galactic nucleus, and the Trapezium, indicating that cluster dynamics will significantly alter the initial distribution function $f_0$ of wide Jupiter-mass binaries.}
    \label{fig:timescale}
\end{figure*}

Fig.~\ref{fig:timescale} shows some examples of the ionization time $\tau$ for different values of $n$ and $\sigma_v$, for Jupiter-mass objects with $m_1 = m_2 = 10^{-3} M_\odot$, and flyby stars of mass $m_3 = 1 M_\odot$. The solid green lines in each panel indicate 1~Myr,  while the red lines mark the Hubble time $\sim 14$~Gyr. The left panel of the figure shows that Jupiter-mass binaries with tight semi-major axis of 1~AU are stable up to the Hubble time in typical open clusters and the Galactic disk, and stable up to 1~Gyr in typical globular clusters, the Galactic nucleus, and the Trapezium cluster. However, as the binary size increases to 100 AU, the ionization time for those binaries is close to 1~Myr in globular clusters, the Galactic nucleus, and Trapezium. Therefore, for wide Jupiter-mass binaries, the initial distribution function $f_0$ will be significantly modified by cluster dynamics.

\subsection{Diffusion model asymptotic solutions}\label{sec:diff}
To obtain the distribution function for extremely soft binaries, we solve Eq.~\ref{eq:f} using the given $\mathcal{C}$, $\mathcal{D}$, and $\mathcal{T}$. For the \textit{diffusion} model where $\mathcal{C}(t>0)=0$, we normalize Eq.~\ref{eq:f} by multiplying by $\tau_c = \tau(\epsilon_c = \frac{Gm_1m_2}{2a_c})$ on both sides, resulting in the dimensionless equation
\begin{eqnarray}
    \frac{\partial f(\hat\epsilon, \hat t)}{\partial \hat t} &=& -\frac{f(\hat\epsilon, \hat t)}{\hat\epsilon} + \frac{2}{3} \int_0^\infty \frac{f(\hat\epsilon^\prime, \hat t) d\hat\epsilon^\prime}{\hat\epsilon^{\prime 1/3} |\hat\epsilon^\prime - \hat\epsilon|^{5/3}} \nonumber \label{eq:f diffusion}   \\
    &&- \frac{2}{3} f(\hat\epsilon, \hat t) \int_0^\infty \frac{d\hat\epsilon^\prime}{\hat\epsilon^{ 1/3} |\hat\epsilon - \hat\epsilon^\prime|^{5/3}}\\
    f(\hat\epsilon, \hat t = 0) &=& f_0(\hat\epsilon) \propto \hat \epsilon^{-\beta_0}\,, 
\end{eqnarray}
where $\hat\epsilon = \epsilon/\epsilon_c$ and $\hat t = t/\tau_c$. Note that the solution of Eq.~\ref{eq:f diffusion} can be rescaled based on parameters such as $n$, $m_{12}$, $m_3$, $\sigma_v$, and $a_{12}$ from Eq.~\ref{eq:tau}.

Although there is no equilibrium solution for $f(\hat\epsilon)$ as previously discussed, we can show that $f(\hat\epsilon,\hat t)$ can reach an asymptotic solution with a constant power-law index $\beta_\infty$. Assuming the solution follows the form $f(\hat \epsilon, \hat t) \propto \hat\epsilon^{-\beta(\hat t)}$, and substituting it into Eq.~\ref{eq:f diffusion}, we obtain\footnote{See Appendix for details.}:

\begin{eqnarray}
     &&g(\beta)=\frac{\partial \beta}{\partial \hat t}=\\
     &&\left\{
     \begin{aligned}
         &\rightarrow+\infty,  &\beta\leq -1&\\
         &\frac{1}{\beta}\Gamma\left(\frac{1}{3}\right)\left(\frac{\Gamma(-\beta+\frac{2}{3})}{\Gamma(-\beta)} + \frac{\Gamma(1+\beta)}{\Gamma(\frac{1}{3}+\beta)} \right), &-1 < \beta < 2/3&\\
         &\rightarrow-\infty, &2/3 \leq \beta&
\end{aligned}\nonumber
     \right.
\end{eqnarray}
The asymptotic value $\beta_\infty$ can be obtained when
\begin{eqnarray}
    \frac{\partial \beta}{\partial \hat t} &=& 0\label{eq:diffusion eqn}\\
    \frac{\partial^2 \beta}{\partial \hat t^2} &\neq& 0 \,,
\end{eqnarray}

\begin{figure}
    \includegraphics[width=\columnwidth]{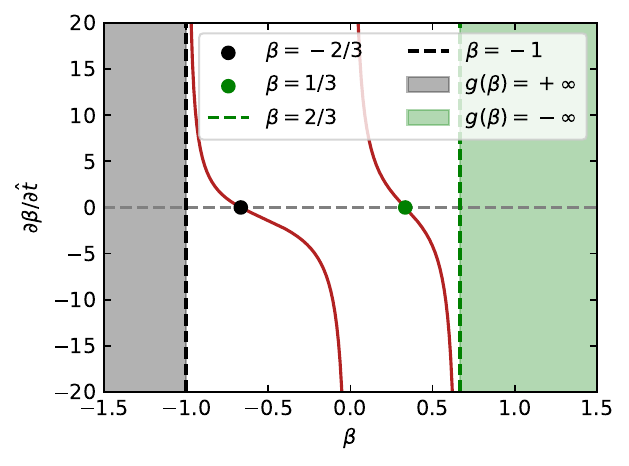}\\
    \includegraphics[width=\columnwidth]{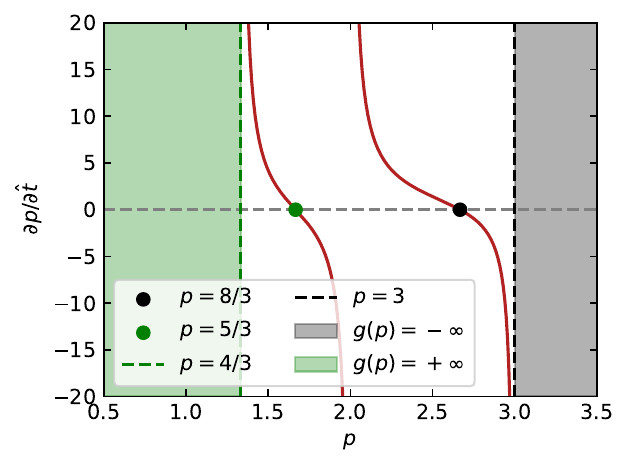}
    \caption{The upper panel shows $g(\beta)=\partial\beta/\partial \hat t$ as a function of $\beta$ for diffusion model. It indicates two solutions for $\partial\beta/\partial \hat t=0$: $\beta_\infty = -2/3$ and $1/3$. For $\beta < -1$, $\partial\beta/\partial \hat t$ is effectively $+\infty$, while for $\beta > 2/3$, $\partial\beta/\partial \hat t$ is effectively $-\infty$. The bottom panel shows the corresponding $\partial p/\partial \hat t = - \partial \beta/\partial \hat t$ as a function of $p$.}
    \label{fig:p}
\end{figure}
The upper panel of Fig.~\ref{fig:p} shows $g(\beta)=\partial\beta/\partial \hat t$ as a function of $\beta$. It indicates that there are two solutions for $\partial\beta/\partial \hat t=0$, $\beta_\infty = -2/3$ and $1/3$. For $\beta < -1$, $\partial\beta/\partial \hat t$ is effectively $+\infty$, causing $\beta$ to rapidly evolve towards $-1$ and then increase to the stable value of $-2/3$ on the timescale of $\tau$. For $\beta> 0$, $\beta$ will eventually stabilize at $-1/3$. This indicates two asymptotic power-law solutions for $f(\hat \epsilon)$:

\begin{eqnarray}
    f(\hat \epsilon) &\propto& \hat\epsilon^{2/3}\\
    f(\hat \epsilon) &\propto& \hat\epsilon^{-1/3}\,.
\end{eqnarray}
Recasting it into a more straightforward form, there are two asymptotic power-law solutions for $f(\hat a)$:
\begin{eqnarray}
    f(\hat a) &\propto& \hat a^{-8/3}\\
    f(\hat a) &\propto& \hat a^{-5/3}\,,
\end{eqnarray}
where $\hat a=a/a_c$. The bottom panel of Fig.~\ref{fig:p} shows the corresponding $\partial p/\partial \hat t = - \partial \beta/\partial\hat t$ as a function of $p$.

We also solve Eq.~\ref{eq:f diffusion} (diffusion model) numerically in the range of $\hat\epsilon\in[10^{-2},10^{3}]$ (which corresponds to $\hat a\in[10^{-3}, 10^{2}]$) from $\hat t= $ 0 to $10$ for different initial $p_0$ = [-1, 0, 1, 2, 3, $\infty$]. We show the results in the form of $f(\hat a,\hat t)$ to better connect with binary sizes.

\begin{figure*}
    \includegraphics[width=2\columnwidth]{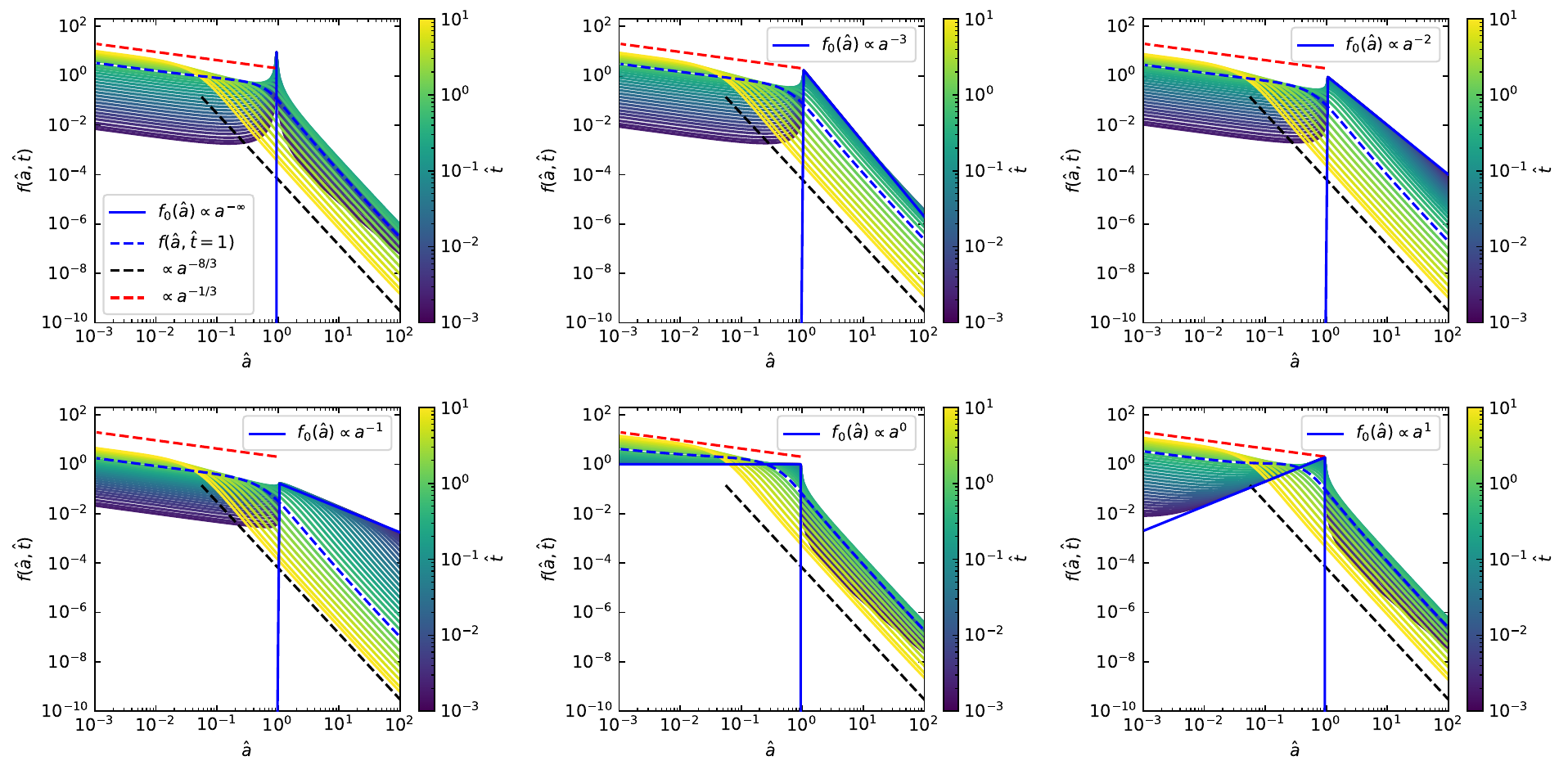}
    \caption{Evolution of $f(\hat a)$ for different initial power-law populations. The initial population is shown with a solid blue line in each panel, while lines of varying colors depict the time evolution of $f(\hat a)$. At $t=C\tau_c$ (with $C$ being a free parameter), binaries with $\hat a > 1/C$ evolve to an asymptotic power-law distribution of $\hat a^{-8/3}$, regardless of the initial populations. This distribution extends to $\hat a=1$ when $\hat t = 1$ (dashed blue line) and to $\hat a = 0.1$ when $\hat t = 10$. }
    \label{fig:eg}
\end{figure*}

Fig.~\ref{fig:eg} shows the evolution of $f(\hat a)$ for different initial power-law populations. In each panel, the initial population is indicated with a solid blue line, and lines with different colors show the time evolution of $f(\hat a)$. It is shown that at $t=C\tau_c$, where $C$ is a free parameter, all binaries with $\hat a > 1/C$ will evolve to an asymptotic population $\propto \hat a^{-8/3}$, independent of the initial populations. The $-8/3$ power-law extends to $\hat a=1$ when $\hat t= 1$ (dashed blue line in each panel) and extends to $\hat a=0.1$ when $\hat t= 10$. This asymptotic power-law is consistent with the solution obtained from Eq.~\ref{eq:diffusion eqn}.

One may wonder why solutions for $p_0<2$ do not converge to $a^{-5/3}$ as indicated by the bottom panel of Fig.~\ref{fig:p}. The reason is that there are no binaries with infinitely large semi-major axes, so the population must be truncated at some large $\hat a$. Therefore, at the far right end of $\hat a$, $f(\hat a\rightarrow\infty)$ is effectively 0, independent of the initial $p_0$. Hence, for the large $\hat a$ end, the effective power-law index is $p=\infty$, corresponding to the red region in the bottom panel of Fig.~\ref{fig:p}. Therefore, regardless of the shape of the initial binary population, all the binaries evolve to $\propto \hat a^{-8/3}$.

\begin{figure}
    \includegraphics[width=\columnwidth]{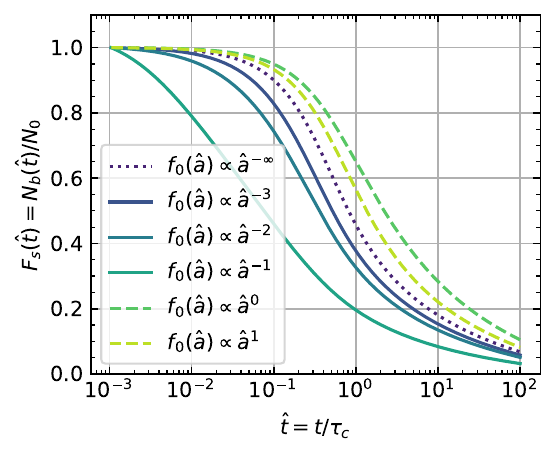}
    \caption{Binary survival fraction $N_b/N_0$ as a function of normalized time $t/\tau_c$ for different initial power-law indices $p_0$. Here, $N_0$ is the total number of binaries at $t=0$, and $N_b(\hat t)$ is the total number of binaries at time $\hat t$. The surviving binaries follow the distribution function indicated by the dashed blue line in the first panel of Fig.~\ref{fig:eg}.}
    \label{fig:frac}
\end{figure}

Of particular interest is the binary survival fraction $N_b/N_0$, where $N_0$ is the total number of binaries at $t=0$ and
\begin{eqnarray}
    N_b(\hat t) = V\int_0^\infty f(\hat a,\hat t) d\hat a
\end{eqnarray}
is the total number of binaries at time $\hat t$, and $N_0=N_b(\hat t =0)$ is the total number of binaries generated at $\hat t=0$. Note that the total number of ionized binaries is
\begin{eqnarray}
    N_i(\hat t) = V\int_{-\infty}^0 f(\hat a,\hat t) d\hat a
\end{eqnarray}
and $N_0 = N_b+N_i$. Fig.~\ref{fig:frac} shows the survival fraction $F_s(\hat t)$ for the \textit{diffusion} model with different $p_0$ as a function of the normalized time $t/\tau_c$. For the case $p=\infty$, where binaries are generated with the same size $a_c$, it is shown that after the ionization time $\tau_c$, approximately 50\% of the binaries will be ionized. The surviving binaries follow the distribution function indicated by the dashed blue line in the first panel of Fig.~\ref{fig:eg}. For shallower initial distributions of $\hat a$ with $p>0$ with $a_c=a_{\rm min}$, the shallower the distribution, the fewer binaries survive after $\tau_c$, as more binaries are initially populated at larger $\hat a$. For $p<0$ with {$a_c = a_{\rm max}$}, since there are no binaries populated at the large $\hat a$ end, as indicated by the solid blue line in the last two panels of Fig.~\ref{fig:eg}, the survival fraction decreases even more slowly.

\subsection{Injection model asymptotic solutions}\label{sec:inj}
For the \textit{injection} model with continuous binary creation, Eq.~\ref{eq:f} can be written in dimensionless form as
\begin{eqnarray}
    \frac{\partial f(\hat\epsilon, \hat t)}{\partial \hat t} &=& \mathcal{\hat C}(\hat\epsilon)  -\frac{f(\hat\epsilon, \hat t)}{\hat\epsilon} + \frac{2}{3} \int_0^\infty \frac{f(\hat\epsilon^\prime, \hat t) d\hat\epsilon^\prime}{\hat\epsilon^{\prime 1/3} |\hat\epsilon^\prime - \hat\epsilon|^{5/3}} \nonumber \label{eq: inj}\\
    &&- \frac{2}{3} f(\hat\epsilon, \hat t) \int_0^\infty \frac{d\hat\epsilon^\prime}{\hat\epsilon^{ 1/3} |\hat\epsilon - \hat\epsilon^\prime|^{5/3}}   \\
    f_0(\hat\epsilon) &=& 0\\
\mathcal{\hat C}(\hat\epsilon)&=&G \hat\epsilon^{-\beta_0}\label{eq:eq f inj}
\end{eqnarray}
where $G$ is a normalization factor that ensures $\mathcal{R}\tau_c=1$ as given by Eq.~\ref{eq:g rate}.

Using a similar approach, we can solve for the asymptotic $\beta_\infty$ from $\frac{\partial \beta}{\partial \hat t}=0$, which yields
\begin{eqnarray}
     0=\frac{\partial \beta}{\partial \hat t} = g(\beta) - u(\beta,\beta_0,\hat \epsilon) \\
      u(\beta,\beta_0,\hat \epsilon) = \hat \epsilon^{\beta - \beta_0 +1}/\beta \label{eq:injection b}
\end{eqnarray}
Unlike for the \textit{diffusion} model, where $\partial \beta/\partial\hat t=0$ is independent of $\hat \epsilon$, Eq.~\ref{eq:injection b} is $\hat \epsilon$ dependent. Since $u < 0$ if $\beta < 0$, and $u > 0$ if $\beta > 0$, as shown in the upper panel of Fig.~\ref{fig:p}, the left root of $g = u$ should range in (-2/3, 0) ($p_\infty \in (2, 8/3)$), and the right root of $g = u$ should range in (0, 1/3) ($p_\infty \in (5/3, 2)$) depending on the value of $u$. The asymptotic solution at large $\hat a$ corresponds to the case $\hat \epsilon \ll 1$. At $\hat t = 0$, $\beta = \beta_0$, thus $u \sim \hat{\epsilon}/\beta_0 \ll 1$. Therefore, at the small $\hat \epsilon$ (large $\hat a$) end, the system evolves like Eq.~\ref{eq:f diffusion} initially. 

\begin{figure}
    \includegraphics[width=\columnwidth]{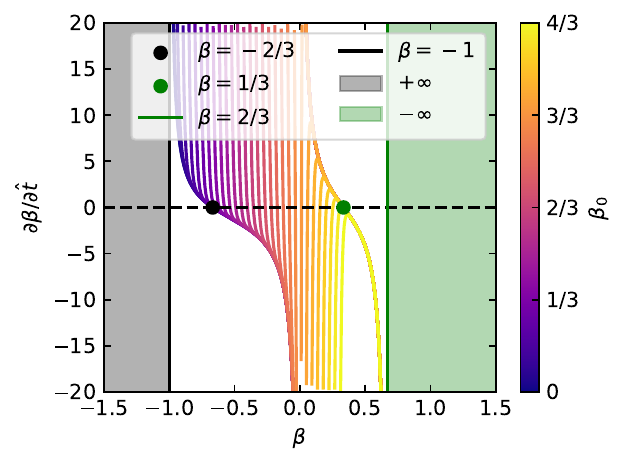}\\
    \includegraphics[width=\columnwidth]{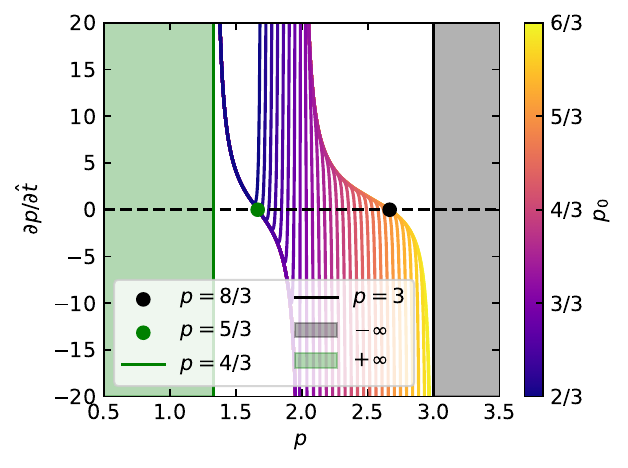}
    \caption{The upper panel shows $\partial\beta/\partial \hat t$ of the injection model as a function of $\beta$, for $\beta_0$ in the range between 0 to 2. Depending on the value of $\beta_0$ the solution for $\partial\beta/\partial \hat t=0$ ranges from -2/3 to 1/3. The bottom panel shows the corresponding $\partial p/\partial \hat t = - \partial \beta/\partial \hat t$ as a function of $p$. The plots are made with $\hat \epsilon=10^{-16}$.}
    \label{fig:u-p}
\end{figure}

1) For $\beta_0 < -2/3$, from the upper panel of Fig.~\ref{fig:p}, we see that $\frac{\partial \beta}{\partial \hat t} > 0$, causing $\beta - \beta_0 + 1$ to increase, leading to smaller $u$, making the system evolve more like Eq.~\ref{eq:f diffusion}. Thus, an asymptotic $\beta_\infty = -2/3$ will be reached.

2) For $-2/3 < \beta_0 < 0$, $\frac{\partial \beta}{\partial \hat t} < 0$, making $\beta - \beta_0 + 1$ smaller. However, even when $\beta$ reaches $\beta_\infty = -2/3$, $\beta - \beta_0 + 1$ will still be positive, thus $u$ will remain $\ll 1$. Therefore, for $\beta_0 < 0$, the system will behave like the \textit{diffusion} model and reach the asymptotic solution with $\beta_\infty = -2/3$.

3) For $0 < \beta_0 < 4/3$, as shown in the upper panel of Fig.~\ref{fig:u-p}, the solution of $\partial \beta/\partial \hat t = 0$ lays in the range between -2/3 and 1/3.

4) For $\beta_0 > 4/3$, as indicated by the upper panel of Fig.~\ref{fig:u-p}, there's no solution for $\partial \beta/\partial \hat t = 0$, while the maximum value of $\partial \beta/\partial \hat t <0$

Translating to $\hat a$, we get the following asymptotic solutions:
\begin{eqnarray}
    f(\hat a) \propto \left\{
    \begin{aligned}
        &\hat a^{-8/3},& \quad  p_0 \geq 2 \\
        &\hat a^{(-8/3, -5/3)}& \quad   2/3 < p_0 < 2 \\
        &\hat a^{-5/3},& \quad p_0 \leq 2/3
    \end{aligned}
    \right.
\end{eqnarray}

\begin{figure*}
\includegraphics[width=2\columnwidth]{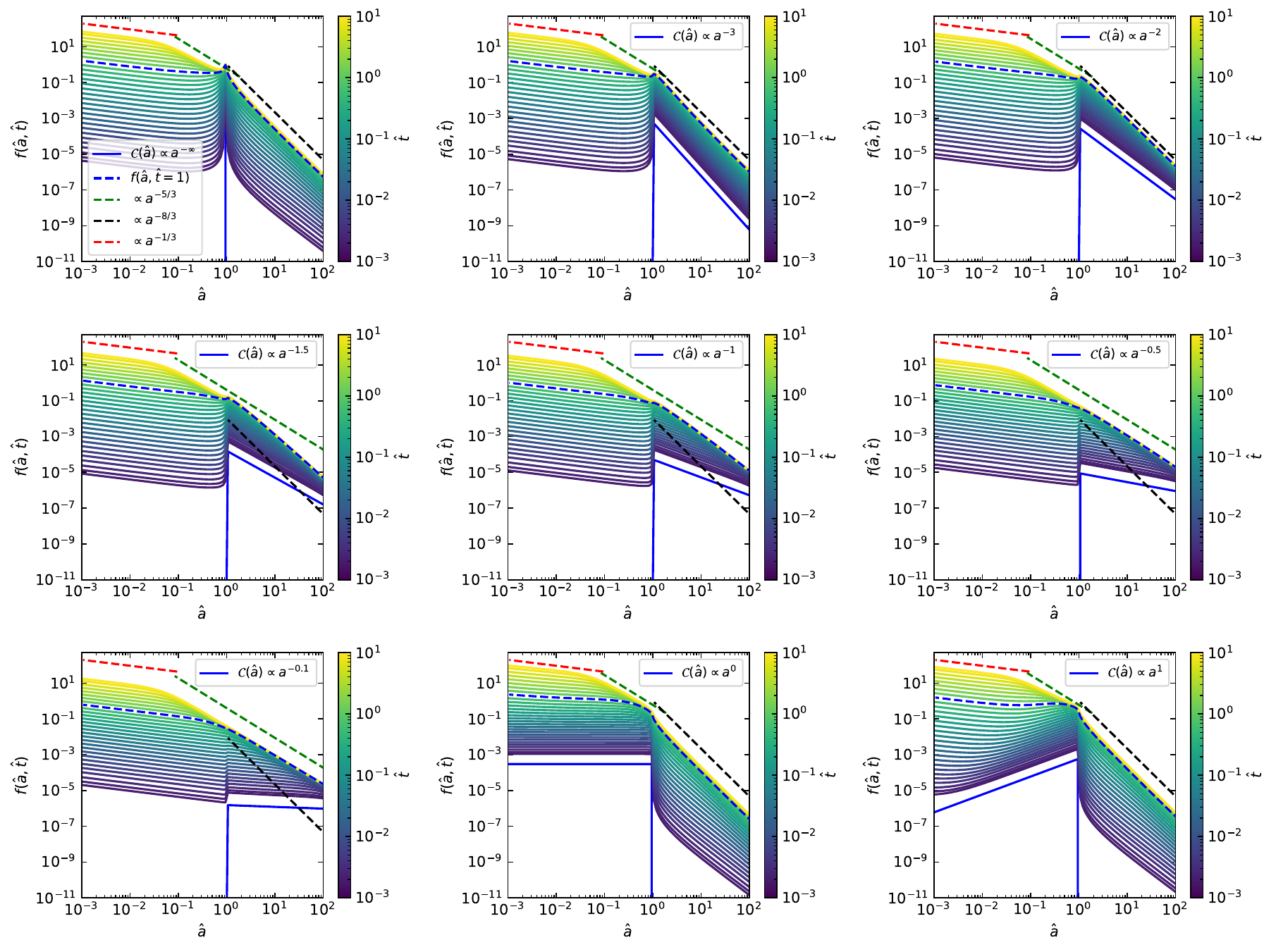}
    \caption{Numerical results for the distribution function $f(\hat a)$ as a function of $\hat a$ for different initial power-law indices $p_0$ based on Eq.~\ref{eq: inj}. The upper panels show that for $\hat t \rightarrow 1$, $\hat a > 1$ evolves to $\hat a^{-8/3}$. The middle panels indicate that for $0 < p_0 < 2$, the asymptotic $p_\infty$ ranges between 8/3 and 5/3. The lower panels show cases with $p_0 > 0$ where the effective $p_0 = \infty$ at large $\hat a$, leading to $p_\infty = 8/3$. For $\hat t > 1$ and $\hat a < 1$, the distribution evolves from $f(\hat a, \hat t = 1)$ (dashed blue lines) toward $p_\infty = 5/3$.}
    \label{fig:eg-c}
\end{figure*}

Fig.~\ref{fig:eg-c} shows the numerical results in terms of $\hat a$ for Eq.~\ref{eq: inj} for different $p_0$. As shown in the upper three panels, for $\hat t \rightarrow 1$, the distribution function with $\hat a > 1$ evolves to asymptotic $\propto \hat a^{-8/3}$. The middle three panels show the result with $0 < p_0 < 2$, indicating that the asymptotic $p_\infty$ ranges between 8/3 and 5/3. The last two panels show the cases with $p_0 > 0$. However, in the large $\hat a$ end, there are no injections, thus the effective $p_0$ is $\infty$, leading to an asymptotic $p_\infty = 8/3$. The cases with $p_0 > 0$ indeed appear when $\hat t > 1$ in the region of $\hat a < 1$. The power-law distribution in the $\hat a < 1$ region starts to build up from the distribution function $f(\hat a, \hat t = 1)$ as shown in dashed blue lines in each panel. One can see that the left side power-law index of those dashed blue lines is $p_0 < 1/3$. Thus, after $\hat t > 1$, the distribution function in the range $\hat a < 1$ evolves toward $p_\infty = 5/3$.

\begin{figure}
    \includegraphics[width=\columnwidth]{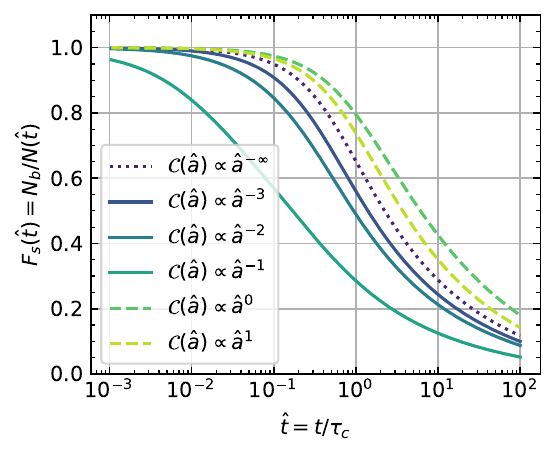}
    \caption{Binary survival fraction as a function of normalized time $\hat t$ for different injection power-law indices $p$. Steeper positive $p$ values ionize more slowly due to fewer binaries at large $\hat a$. The injection model is more efficient at maintaining binary populations compared to the diffusion model.}
    \label{fig:frac-c}
\end{figure}

Fig.~\ref{fig:frac-c} shows the survival fraction $F_s(\hat t)$ as a function of the normalized time $\hat t$ for different injection power-law indices $p$. Similarly to the diffusion model, binary populations with steeper positive $p$ are ionized more slowly due to the less populated binaries at the large $\hat a$ end. Unlike the diffusion model, the injection model is better at maintaining binaries. For the case $p = \infty$, i.e., binaries are generated all with the same size $a_c$, after the ionization time $\tau_c$, it is shown that approximately 30\% of the binaries are ionized. The surviving binaries follow the distribution function indicated by the dashed blue line in the first panel of Fig.~\ref{fig:eg-c}.

\section{Application to the jumbo candidates in the trapezium cluster}
In this section we apply our generalized results to the recent report of candidate Jupiter Mass Binary Objects (JuMBOs), and impose constraints on their possible formation mechanisms, if their existence is confirmed.

\citet{Pearson2023} reported the discovery of approximately 540 planet-mass objects (PMOs) in the Trapezium cluster of the Orion Nebula. Among these, 40 systems were identified as visual binaries with semi-major axes ranging from 25 AU to 390 AU. Binaries with semi-major axes smaller than 25 AU were not resolved to the instrumental resolution.
\begin{figure}
    \includegraphics[width=\columnwidth]{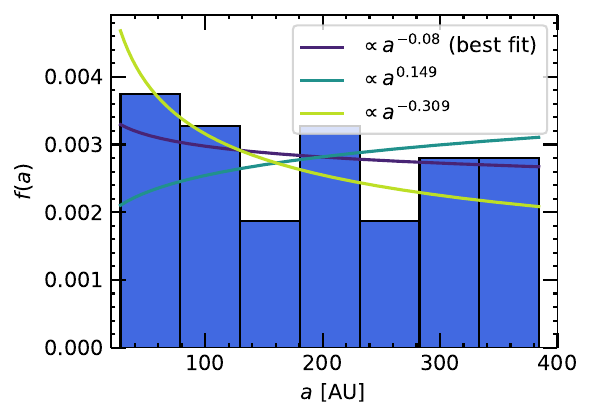}
    \caption{The semi-major axis distribution of observed Jupiter Mass Binary Objects (JuMBOs) in the Trapezium cluster of the Orion Nebula. The distribution appears effectively flat, indicated by $p_{\rm JuMBO} \sim 0$.}
    \label{fig:a-obs}
\end{figure}
Fig.~\ref{fig:a-obs} shows the semi-major axis distribution of the observed JuMBOs, which is effectively flat. Thus, we denote the index of the distribution by  $p_{\rm JuMBO} \sim 0$. Based on these observations, we can constrain JuMBO formation mechanisms using our soft binary evolution model in star clusters. 

We use the following four observational constraints:

\begin{itemize}
\item 1. The total number of JuMBOs observed in the 25-390 AU range is approximately 40, with an age of $t_{\rm obs} \sim 1$ Myr.
\item 2. The total number of PMOs is roughly 540.
\item 3. The JuMBO to PMO ratio is $\sim 9\%.$
\item 4. The  power law index of the semi-major axis distribution
is $\sim 0$.
\end{itemize}

To apply these constraints, we rescale our results for $f(\hat a,\hat t)$ obtained from solving Eqs.~\ref{eq:f diffusion} and \ref{eq:eq f inj}, shown in Figs.~\ref{fig:eg} and \ref{fig:eg-c}, using Eq.~\ref{eq:tau} with the following parameters: $m_1 = m_2 = 1 M_J$,  $m_3 = 1 M_\odot$, $n=5\times10^4$ pc$^{-3}$, and $\sigma_v=$ 2 km/s for the cluster properties of the Trapezium. The cluster properties are taken from \cite{Hillebrand1998}. According to this work, there are roughly 3000 stars in the Orion Nebula. We aim at constraining the most populated JuMBO size $ a_c$ at $t=0$, the initial semi-major axis distribution power law index $p_0$, and the total number of JuMBOs formed over 1~Myr, $N_0$ (hence the JuMBO formation rate $N_0$/1 Myr).

The total number of JuMBOs within the 25-390 AU range, $N_{\rm J, obs}$, is given by 
\begin{eqnarray}
    N_{\rm JuMBO, obs} = N_0 \int_{25 \rm AU}^{390 \rm AU} f(a, t_{\rm obs}) da\,.
\end{eqnarray}
The total number of PMOs, $N_{\rm PMO, obs}$, is calculated via
\begin{eqnarray}
    N_{\rm PMO, obs} &=& N_0 \bigg(\int_0^{25 \rm AU} f(a, t_{\rm obs}) da + 2\int_{25 \rm AU}^{\infty} f(a, t_{\rm obs}) da \nonumber\\
    &+& 2\int_{-\infty}^{0} f(a, t_{\rm obs}) da\bigg) + N_{\rm SFP}\,.
\end{eqnarray}
The first term represents tight JuMBOs with $a<$ 25 AU identified as single PMOs, the second term represents observed JuMBOs and super wide JuMBOs identified as single PMOs, contributing 2 PMOs per system, the third term represents ionized JuMBOs, and the last term represents the total number of single free-floating planets (SFP) formed from planetary system instability or ejection, excluding those from JuMBO ionizations. Since $N_{\rm PMO, obs} > N_{\rm SFP}$ and there are roughly 540 PMOs, $N_{\rm SFP}$ should not exceed 540, leading to an upper limit of Jupiter mass free-floating planet (FFP) formation rate (excluding JuMBO ionization) of $\sim 540$ Myr$^{-1}$. Assuming all these FFPs come from planetary systems with host stars, this rate translates to $\sim 540/3000 \sim 0.18$ Myr$^{-1}$ per planetary system.

We can also constrain the lower limit of $N_{\rm SFP}$. In the Trapezium, stellar flybys eject Jupiter-mass planets from planetary systems, leading to the formation of PMOs. The single planet ejection cross section from flyby is given by Eq.~\ref{eq:hut} (equal-mass case where $m_{12} \sim m_3$), therefore:
\begin{eqnarray}
    N_{\rm SFP, min}\sim n \sigma_v t_{\rm obs}\pi\left(\frac{Gm_{3}}{\sigma_v^2}\right)a_{\rm Jup}\,.
\end{eqnarray}
 Plugging in the numerical values of  $n$ and $\sigma_v$ specific to Trapezium, and assuming Jupiter-mass planets are most populated at $a_{\rm Jup} \sim \mathcal{O}(10)$~AU (as for our Solar system), leads to $N_{\rm SFP, min} \sim 50$. $N_{\rm SFP, min}$ should be higher if other direct ejection mechanisms \citep{Rasio1996,Raymond2010,Chen2024} are considered. The JuMBO to PMO ratio is given by
\begin{eqnarray*}
     F_{\rm JuMBO} = N_{\rm JuMBO}/N_{\rm PMO, obs}\,.
 \end{eqnarray*}

The actual formation mechanisms of the reported JuMBO candidates are still unknown. The two most straightforward mechanisms are in-situ formation, where JuMBOs form like binary stars \citep{Portegies2023}, and ejection, where JuMBOs form via ejection from planetary systems \citep{Wang2024, Lazzoni2024,Yu2024}. For in-situ formation, where binaries form roughly uniformly in log scale of $a$, i.e., $\mathcal{C}(a)\propto dN/d\log(a) = \text{const}$ \citep{Connelley2008, Chen2013}, we have: 
\begin{eqnarray}
     \mathcal{C}_{\rm in}(a)\propto a^{-1}\,,
 \end{eqnarray}
i.e., $p_0 = 1$. For the ejection scenario, since the formation cross section is $\sigma_{\rm ej}\propto a_{\rm Jup}^2$, if we assume $a_{\rm Jup}$ is uniformly distributed, then:
\begin{eqnarray}
     \mathcal{C}_{\rm ej}(a)\propto \frac{dN}{da} \propto \frac{d\sigma_{\rm ej}}{da}\propto a^{1}\,, \end{eqnarray}
i.e., $p_0 = -1$.

We calculate $N_{\rm JuMBO, obs}$, $N_{\rm PMO, obs}$, $F_{\rm JuMBO}$, and the power law index of the
semi-major axis distribution  for different $p_0$, $N_0$, $a_c$, and $N_{\rm SFP}$, for both the diffusion model and the injection model described in Section~\ref{sec:diff} and \ref{sec:inj}.

\begin{figure*}
    \includegraphics[width=2\columnwidth]{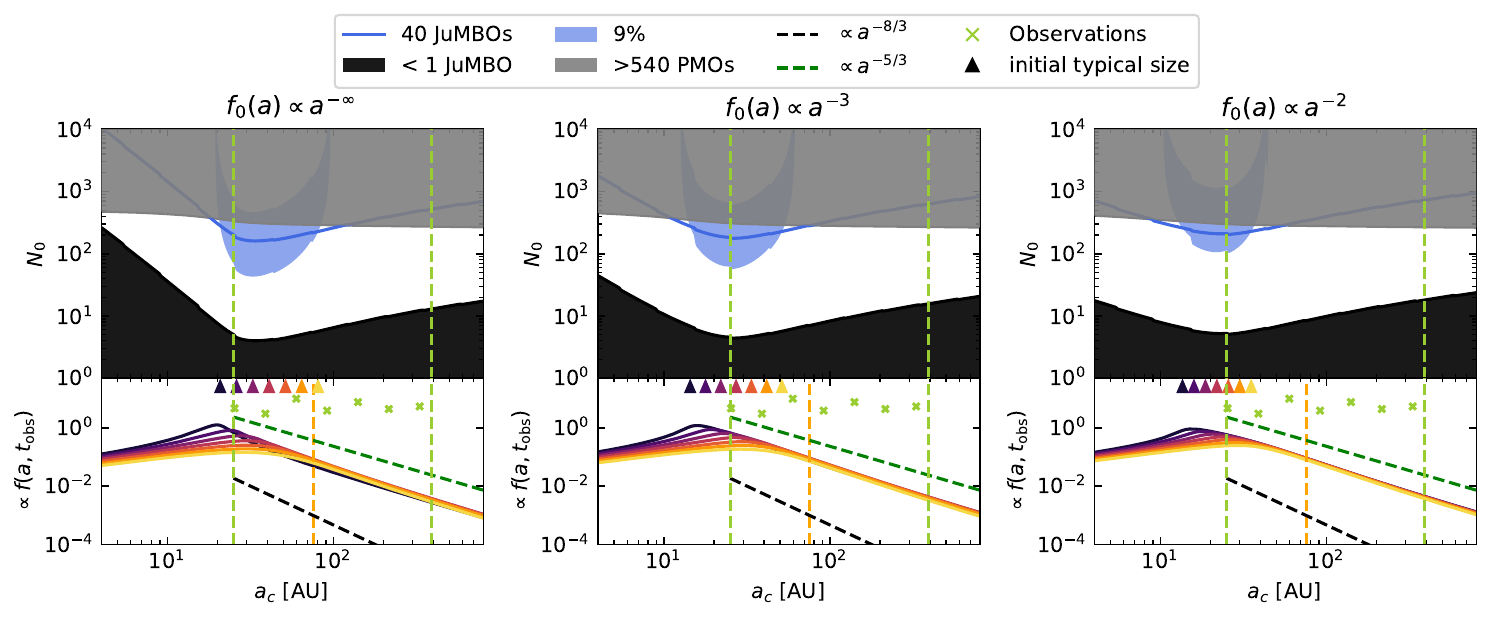}
    \includegraphics[width=2\columnwidth]{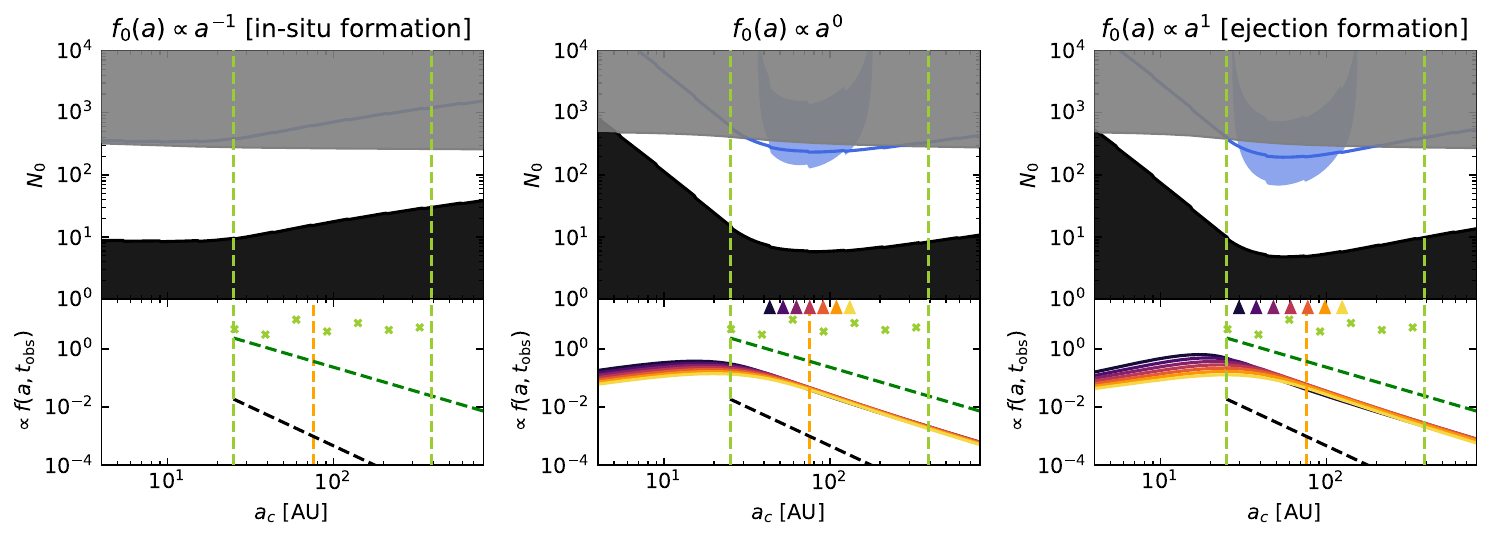}
    \caption{Constraints on the diffusion model for JuMBO formation. Each panel shows the required initial number of JuMBOs, $N_0$, to produce the observed 40 JuMBOs for different power law indices $p_0$. The solid blue contour line indicates where $N_{\rm JuMBO, obs} = 40$, while the black contour line represents 1 JuMBO. Black shaded regions denote parameter spaces where JuMBO formation is not possible. The gray solid line shows the upper limit of $N_0$ corresponding to $N_{\rm PMO, obs} = 540$ with $N_{\rm SFP} = 0$. Blue shaded regions indicate where the JuMBO fraction $F_{\rm JuMBO}$ is 9\%, with the lower boundary for $N_{\rm SFP} = 50$ and the upper boundary for $N_{\rm SFP} = 540$. Orange dashed line marks the semi-major axis $a_d \sim 80$ AU, where the ionization timescale is 1 Myr. Bottom subpanels display the rescaled (for better visualization) semi-major axis distribution of JuMBOs at $t_{\rm obs} = 1$ Myr for different constrained typical JuMBO size $a_c$ indicated by colored triangles, with observational data represented by green crosses.}
    \label{fig:jumbo-obs-diff}
\end{figure*}

Fig.~\ref{fig:jumbo-obs-diff} shows our results for the diffusion model (all JuMBOs formed at $t=0$) for different $p_0$. { \textit{The total number of observed JuMBOs constrains $N_0$ and $a_c$ along the blue solid lines:}} The number $N_{\rm JuMBO, obs}$ is shown in solid contour lines, where the solid blue line indicates $N_{\rm JuMBO, obs}=40$, and the { black region indicates $< 1$ JuMBOs.} For all $p_0$, if $a_c \ll 25$ AU, very few JuMBOs have $a$ in the 25-390 AU range, indicated by the two vertical dashed green lines. As $a_c$ decreases, a larger $N_0$ is required to produce 40 JuMBOs in the observed range.

\textit{ The number of PMOs constrains $N_0$ below the grey regions:} The gray solid line shows the upper limit of $N_0$, corresponding to $N_{\rm PMO, obs}=540$ and $N_{\rm SFP}=50$. { Since the real $N_{\rm SFP}$ should be higher than 50, the allowable $N_0$ from this $N_{\rm PMO,obs}$ constraint should be lower than the grey region.} A large $N_0$ would lead to more $N_{\rm PMO, obs}$ from JuMBO ionizations, thus overproducing $N_{\rm PMO, obs}$ compared to observations. To produce 40 observed JuMBOs, the most populated JuMBO formation size $a_c$ is constrained to $\mathcal{O}(10)$ AU for $p_0 \geq 2$ and $\mathcal{O}(100)$ AU for $p_0 < 0$ (for $p_0 < 0$, the initial distribution tail is on the small $a$ side). The case $p_0=1$ is ruled out as the 40 JuMBO contour line is above the gray line $N_{\rm PMO, obs}=540$ limit.

\textit{ The 9\% JuMBO fraction constrains $N_0$ and $a_c$ within the blue shaded regions:} The blue shaded region shows the parameter space where $F_{\rm JuMBO}=9\%$ for  $N_{\rm SFP}$ ranging from 50 to 540. The lower boundary indicates the case where $N_{\rm SFP}=50$, and the upper boundary corresponds to $N_{\rm SFP}=540$. The figure shows that the 9\% $F_{\rm JuMBO}$ and 40 JuMBOs can be achieved simultaneously for most $p_0$ values, but not for $p_0=1$ (in-situ formation).

\textit{ The flat distribution of JuMBO semi-major axis constrains the model parameter $p$:} The dashed orange line indicates the semi-major axis $a_d \sim 80$ AU, where the corresponding ionization timescale $\tau_c$ (Eq.~\ref{eq:tau}) is 1 Myr. As shown in Fig.~\ref{fig:eg} and discussed in Section~\ref{sec:diff}, JuMBOs with $a > a_d$ will gradually evolve towards the asymptotic power law distribution $\propto a^{-8/3}$ or $\propto a^{-5/3}$, which are both 
far from the observed $a^{\sim 0}$ in the $>80$ AU range shown in Fig.~\ref{fig:a-obs}. The bottom subpanel in each panel shows the distribution of the JuMBO semi-major axis at $t_{\rm obs}=$ 1 Myr for different initial typical JuMBO sizes $a_c$ indicated by colored triangles. 
%in the constraint ranges. 
The green crosses represent the semi-major distribution of the observed data. We obtained it by binning the reported data \citep{Pearson2023} so that there are 6 points in each bin. Note that the absolute normalization is meaningless. None of the cases can match the observed $p_{\rm JuMBO}{\sim 0}$, thus ruling out all cases for the diffusion model.

{    While this paper was being written, \cite{Huang2024} performed an independent study on constraining the initial semi-major axis of the JuMBOs for $N_0$ = 1000 and $p_0 = \infty$. They found that the required $a_c$ is roughly $\mathcal{O}(10)$ AU for 9\% JuMBO fraction. Although this is consistent with our results, we wish to emphasize that if the
additional observational constraint  $p_{\rm JuMBO}\sim$0 is considered, none of the models becomes viable.  }

\begin{figure*}
    \includegraphics[width=2\columnwidth]{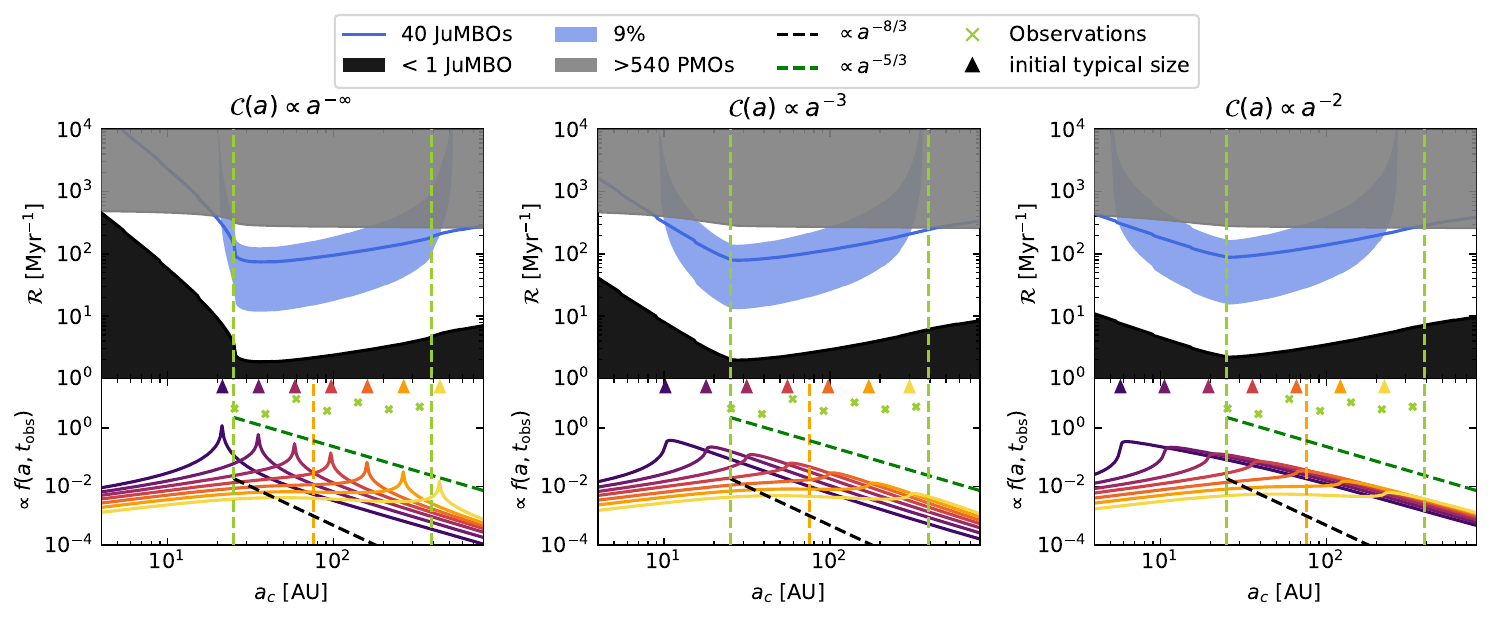}
    \includegraphics[width=2\columnwidth]{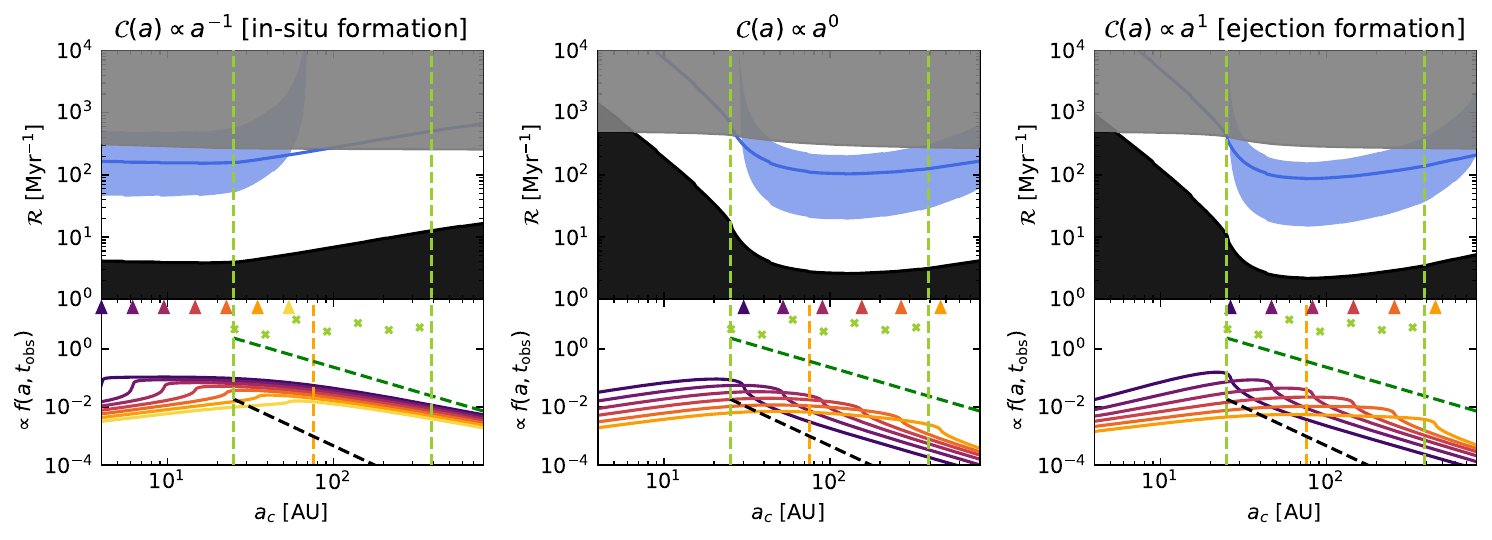}
    \caption{Constraints on the continuous injection model for JuMBO formation. Each panel shows the required $N_0$ to produce the observed 40 JuMBOs for different $p_0$. The solid blue line in each panel marks $N_{\rm JuMBO, obs} = 40$, and the black line marks 1 JuMBO. The orange dashed line indicates the semi-major axis $a_d \sim 80$ AU, where the ionization timescale is 1 Myr. The bottom subpanels display the distribution of JuMBO semi-major axes at $t_{\rm obs} = 1$ Myr for different initial sizes $a_c$ indicated by colored triangles. The green crosses represent the observational data. The injection model can produce a flat distribution of semi-major axes, especially for $p_0 = 0$ and $-1$, but fails for in-situ formation ($p_0 = 1$) and certain other scenarios.}
    \label{fig:jumbo-obs-inj}
\end{figure*}

Fig.~\ref{fig:jumbo-obs-inj} shows the results of the continuous injection model. The injection model requires a lower $N_0=\mathcal{R}t_{\rm obs}$ compared to the diffusion model, since the injection model tends to preserve more JuMBOs. A wider range of $a_c$, from tens to hundreds of AU, is able to reproduce 40 observed JuMBOs with a 9\% JuMBO fraction. The injection model is better at producing a flatter semi-major axis distribution. The flat distribution between 25-390 AU can be achieved for most $p_0$ values, except for the in-situ formation case with a slight decay at the large end. For $p_0=\infty$ (i.e., $\delta$ function injection), if the typical injection $a_c$ is around the maximum size of the observed JuMBOs, a flat distribution below $a_c$ can be established due to binary hardening. However, a peak around $a_c$ shows up here which is not observed in the data. This peak will disappear as $a_c$ increases, but a larger $a_c$ will fail to reproduce the 9\% JuMBO fraction. For $p_0 = 3$, the flat distribution can be achieved as $a_c$ approaches its upper limit $\sim 350$ AU, narrowly passing all tests. However, no known JuMBO formation mechanism follows $\mathcal{C}\propto a^{-3}$. For $p_0 = 2$, the flat distribution fails in the $>$200 AU region. For $p_0=0$ and -1, a flat distribution can be achieved if $a_c\sim 390$ AU, with a rapid decline for $a > a_c$. Therefore, the ejection mechanism and mechanisms with $\mathcal{C}\propto a^0$ can pass all tests. However, as indicated by the solid blue line, if $a_c$ exceeds 390 AU, the required formation rate must be $>$ 120 Myr$^{-1}$ ($>$0.04 Myr$^{-1}$ per planetary system). This requires extremely wide ($>$500 AU) pair Jupiter orbits around each star in the Trapezium, which may be challenging \citep{Wang2024}. In summary, for the injection model, only JuMBO formation mechanisms with $p=3, 0$, and $-1$ (ejection mechanism) can pass all tests, with a JuMBO formation rate $>$ 120 Myr$^{-1}$ in the Trapezium.

If the $N_{\rm JuMBO, obs}$, $N_{\rm PMO, obs}$, $F_{\rm JuMBO}$, and $p_0$ are measured correctly, then to better match the observed flat semi-major axis distribution $\propto a^{\sim0}$ between 25-390 AU, $a_d$ needs to move to the larger side, making the asymptotic distribution $\propto a^{-p_\infty}$ with $p_\infty \in [5/3, 8/3]$ outside the observed semi-major axis range. Due to the scaling of Eq.~\ref{eq:tau}, if the JuMBO mass increases by a factor of $\sim 100$, e.g., if these JuMBOs are rather $10^{-1}M_\odot$ dwarfs, then $a_d$ will move from $\sim 80$ AU to $400$ AU.

\begin{figure*}
    \includegraphics[width=2\columnwidth]{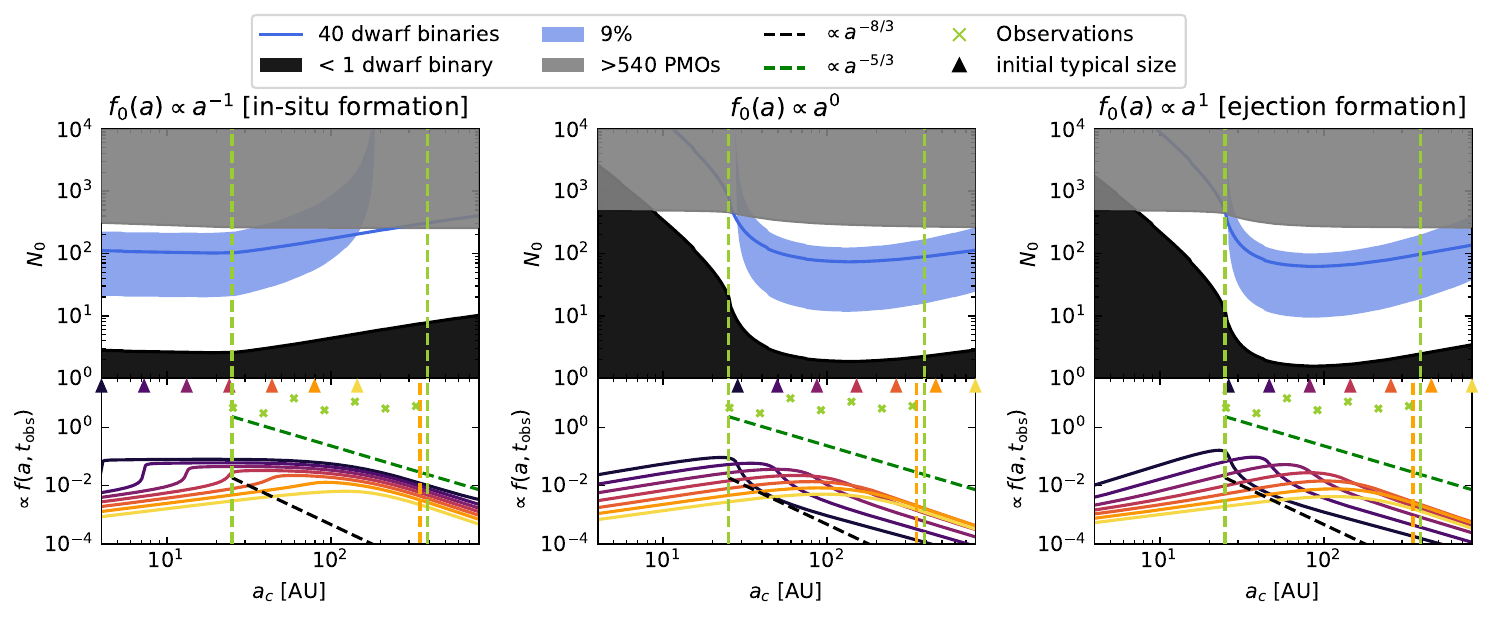}
    \includegraphics[width=2\columnwidth]{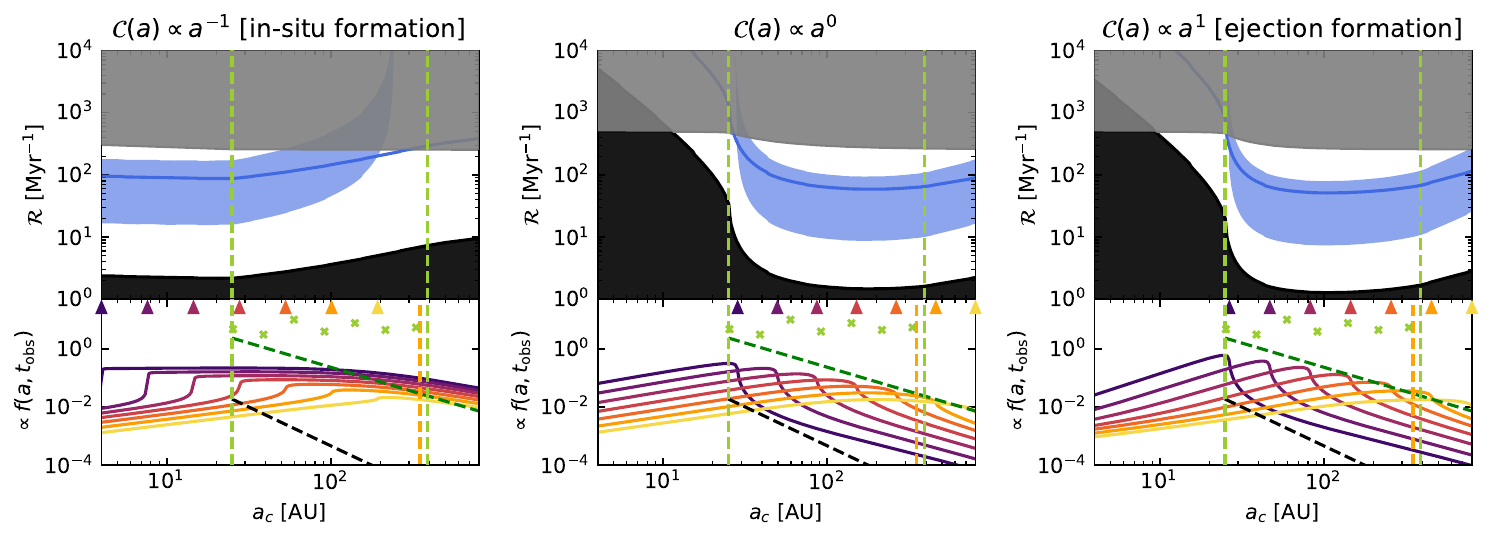}
    \caption{Parameter space constraints for binary brown dwarf formation. The upper three panels show the diffusion model, while the lower three panels show the continuous injection model. For each model, the parameter space is shown for different $p_0$ values. Black shaded regions indicate where binary brown dwarf formation is not possible. Blue shaded regions show where the brown dwarf fraction is 9\%, with varying $N_{\rm SFP}$. The continuous in-situ formation mechanism with $a_c \sim 20$ AU and a formation rate of 100 Myr$^{-1}$ is identified as the most promising scenario.}
    \label{fig:dwarf}
\end{figure*}
As shown in Fig.~\ref{fig:dwarf}, if these JuMBOs were indeed dwarf binaries, a much wider parameter space could establish the flat semi-major axis distribution. The upper three panels show the case for diffusion model, and the bottom three panels show the injection model. For dwarf binaries, although the required binary formation rate is lower than for JuMBOs, the ejection mechanism becomes irrelevant as pair dwarfs are unlikely to form protoplanetary disks, providing no initial condition for the ejection mechanism. The most promising explanation is the continuous in-situ formation mechanism with $a_c\sim 20$ AU and a formation rate of 100 Myr$^{-1}$, which is not difficult to achieve in the Trapezium.

\section{Discussion }

As discussed in the previous section, we found that JuMBO formation mechanisms face significant challenges in meeting all four observational constraints. While the in-situ formation mechanism, with its relatively high formation rate, can explain the 40 JuMBOs, the 540 PMOs, and the 9\% JuMBO fraction, it struggles to account for the observed flat JuMBO semi-major axis distribution. Conversely, the ejection formation mechanism better matches the flat semi-major axis distribution but fails to achieve the required formation rate to account for 40 surviving JuMBOs over 1 Myr.

To address this issue, we demonstrated that loosening the observational constraints, particularly by considering higher measured masses in the brown dwarf range, ameliorates the discrepancy.
This adjustment makes it easier for these binaries to meet all four observational constraints, especially the semi-major axis distribution. Higher mass binaries require a smaller formation rate and are less affected by stellar flybys after their birth, resulting in a flatter semi-major axis distribution within the observed range.

Beyond increasing the binary mass, other approaches to adjusting the measured quantities could better align with the four observational data points. Essentially, we need to either increase the characteristic timescale ($\tau_c$) or shorten the observational time ($t_{\rm obs}$) to reduce $\hat t = t_{\rm obs}/\tau_c$ so that the effects of surrounding stellar flybys on these binaries are weaker. To meet the four observed constraints, $\tau_c$ needs to be increased by a factor of five. From Eq.~\ref{eq:tau}, instead of increasing the binary mass $m_{12}$ by a factor of roughly 100, we could achieve a larger $\tau_c$ by having a lower stellar number density $n$ and a higher velocity dispersion $\sigma_v$. However, given the correlation between $n$ and $\sigma_v$ during cluster evolution and the Trapezium cluster's young age and ongoing core collapse, achieving the required factor of five increase in $\tau_c$ may be difficult.

Another approach is to reduce $t_{\rm obs}$ by a factor of five, implying that these binaries are much younger, around $2 \times 10^5$ years, as proposed by previous studies (e.g., \citealt{Portegies2023}). This would suggest that the measured age of these JuMBOs \citep{Pearson2023} has been overestimated and leads to the conclusion that lower mass binaries formed more slowly than higher mass binary stars in Trapezium.

\section{Summary}
In this study, we have investigated the scattering interactions between extremely soft binaries and single stars within stellar clusters to determine the correct cross section for these interactions. We have provided a detailed framework for understanding the scattering interactions and evolutionary dynamics of extremely soft binaries in stellar clusters, via a combination of analytical and numerical methods. 

Using the impulsive approximation, we find that the cross section for small-mass binaries interacting with a flyby star is smaller than for the cases where the binary and the flyby star have equal masses, deviating significantly from established equal-mass scenarios \citep{Hut1983}. 

For the 
binary semi-major axis distribution function, 
the equilibrium state for equal mass cases has been well studied in the literature. However, for small-mass, extremely soft binaries, the equilibrium state cannot be reached. Instead, we find asymptotic solutions $\propto a^{-p}$ for the distribution function on the timescale of $\tau_c$. Using both diffusion and continuous injection models, we determine that in the diffusion model, the asymptotic solution for the distribution function is proportional to $a^{-8/3}$ at the soft end. In the continuous injection model, where binaries form steadily over time, the asymptotic solutions range from $a^{-8/3}$ to $a^{-5/3}$, depending on the initial power-law index of the injection. Specifically, if the injection power-law index is steeper than $a^{-2}$ or shallower than $a^{0}$, the distribution will evolve to $a^{-8/3}$; otherwise, it will evolve to asymptotic solutions between $a^{-8/3}$ and $a^{-5/3}$.

Applying our theoretical models to the recent discovery of Jupiter-mass binary objects (JuMBOs) in the Trapezium cluster, we have used four observed quantities 
to constrain their formation mechanisms and evolutionary scenarios: (1) the number of observed JuMBOs within the 25-390 AU range is 40, (2) the total number of PMOs in the cluster is 540, (3) the JuMBO to PMO ratio is 9\%, and (4) the power-law index of the observed semi-major axis distribution is roughly 0.

We have found that for the diffusion model, producing the observed 40 JuMBOs within the semi-major axis range of 25-390 AU and a 9\% JuMBO fraction requires a formation size around $\mathcal{O}(10)$~AU for initial power-law indices $p_0 \geq 2$ and $\mathcal{O}(100)$ AU for $p_0 < 0$. However, this model fails to match the observed flat distribution, indicating that the observed JuMBOs are unlikely to be formed within a very short timescale compared to their 1~Myr ages.

The continuous injection model, particularly with an initial power-law index $p = -1$ (ejection mechanism), matches all four observed quantities better. However, this model requires a JuMBO formation rate exceeding 120 Myr$^{-1}$, which is challenging to achieve. Therefore, the most plausible explanation is that the mass of the observed JuMBOs has been underestimated.

If JuMBOs are actually binary brown dwarfs, a continuous in-situ formation mechanism with a typical initial semi-major axis around 20 AU and a formation rate of 100 Myr$^{-1}$ is favored. This scenario is consistent with maintaining a stable binary population and matching the observed distributions, providing a plausible explanation for the observed flat semi-major axis distribution.

\section*{acknowledgments}
Y.W. and Z.Z. acknowledge NASA 80NSSC23M0104 and Nevada Center for Astrophysics for support. Z.Z. acknowledges support from the National Science Foundation under CAREER grant AST-1753168 and support from NASA award 80NSSC22K1413. 
R.P. acknowledges support by NSF award AST-2006839.
%\section{acknowledgments}

%% To help institutions obtain information on the effectiveness of their 
%% telescopes the AAS Journals has created a group of keywords for telescope 
%% facilities.
%
%% Following the acknowledgments section, use the following syntax and the
%% \facility{} or \facilities{} macros to list the keywords of facilities used 
%% in the research for the paper.  Each keyword is check against the master 
%% list during copy editing.  Individual instruments can be provided in 
%% parentheses, after the keyword, but they are not verified.

%\vspace{5mm}
%\facilities{}

%% Similar to \facility{}, there is the optional \software command to allow 
%% authors a place to specify which programs were used during the creation of 
%% the manuscript. Authors should list each code and include either a
%% citation or url to the code inside ()s when available.

\software{{\tt SpaceHub} \citep{Wang2021}; {\tt NumPy} \citep{2020NumPy-Array}.}

%% Appendix material should be preceded with a single \appendix command.
%% There should be a \section command for each appendix. Mark appendix
%% subsections with the same markup you use in the main body of the paper.

%% Each Appendix (indicated with \section) will be lettered A, B, C, etc.
%% The equation counter will reset when it encounters the \appendix
%% command and will number appendix equations (A1), (A2), etc. The
%% Figure and Table counter will not reset.

\appendix

%\begin{widetext}
\begin{eqnarray}
\delta\mathbf{v}_1 &=&\frac{Gm_3}{h_3}\left(
\begin{aligned}
\int_{-\nu_\infty}^{+\nu_\infty}\frac{r_3^2(r_3\cos\nu- r_1\cos\bar{\omega})d\nu}{(r_3^2+r_1^2-2r_3r_1\cos(\nu-\bar{\omega}))^{3/2}}\\
\int_{-\nu_\infty}^{+\nu_\infty}\frac{r_3^2(r_3\sin\nu- r_1\sin\bar{\omega})d\nu}{(r_3^2+r_1^2-2r_3r_1\cos(\nu-\bar{\omega}))^{3/2}}
\end{aligned}
\right)\\
&=&\frac{Gm_3}{bv_\infty}\left(
\begin{aligned}
\int_{-\nu_\infty}^{+\nu_\infty}[\cos\nu + \frac{r_1}{r_3}(3\cos^2\nu\cos\bar{\omega}+\cos\nu\sin\nu\sin\bar{\omega}-\cos\bar{\omega})+\mathcal{O}(\frac{r_1^2}{r_3^2})]d\nu\\
\int_{-\nu_\infty}^{+\nu_\infty}[\sin\nu + \frac{r_1}{r_3}(3\cos\nu\sin\nu\cos\bar{\omega}+\sin^2\nu\sin\bar{\omega}-\sin\bar{\omega})+\mathcal{O}(\frac{r_1^2}{r_3^2})]d\nu\\
\end{aligned}
\right)\\
&=&\frac{Gm_3}{bv_\infty}\left(
\begin{aligned}
\int_{-\nu_\infty}^{+\nu_\infty}[\cos\nu + \frac{r_1}{l_3}\cos\bar{\omega}(1+e_3\cos\nu)(3\cos^2\nu-1)]d\nu\\
\int_{-\nu_\infty}^{+\nu_\infty}[\sin\nu + \frac{r_1}{l_3}\sin\bar{\omega}(1+e_3\cos\nu)(3\sin^2\nu-1)]d\nu\\
\end{aligned}
\right)\\
&=&\frac{Gm_3}{bv_\infty}\left(
\begin{aligned}
&&\frac{2\sqrt{e_3^2-1}}{e_3}+\frac{r_1}{l_3}\cos\bar{\omega}\left((2+\frac{5}{e_3^2})\sqrt{e_3^2-1}+\nu_\infty\right)\\
&&\frac{r_1}{l_3}\sin\bar{\omega}\left(\nu_\infty-\frac{5}{e^2_3}\sqrt{e_3^2-1}\right)
\end{aligned}
\right)\\
\delta\mathbf{v}_2 &=&\frac{Gm_3}{h_3}\left(
\begin{aligned}
\int_{-\nu_\infty}^{+\nu_\infty}\frac{r_3^2(r_3\cos\nu + r_2\cos\bar{\omega})d\nu}{(r_3^2+r_2^2+2r_3r_2\cos(\nu-\bar{\omega}))^{3/2}}\\
\int_{-\nu_\infty}^{+\nu_\infty}\frac{r_3^2(r_3\sin\nu+ r_2\sin\bar{\omega})d\nu}{(r_3^2+r_2^2+2r_3r_2\cos(\nu-\bar{\omega}))^{3/2}}
\end{aligned}
\right)\\
&=&\frac{Gm_3}{bv_\infty}\left(
\begin{aligned}
&&\frac{2\sqrt{e_3^2-1}}{e_3}-\frac{r_2}{l_3}\cos\bar{\omega}\left((2+\frac{5}{e_3^2})\sqrt{e_3^2-1}+\nu_\infty\right)\\
&&-\frac{r_2}{l_3}\sin\bar{\omega}\left(\nu_\infty-\frac{5}{e^2_3}\sqrt{e_3^2-1}\right)
\end{aligned}
\right)\\
\end{eqnarray}

\begin{eqnarray}
    I_1= \int_0^\infty \frac{\hat\epsilon^{\prime -\beta}d\hat\epsilon^\prime}{\hat\epsilon^{\prime 1/3}|\hat\epsilon-\hat\epsilon^\prime|^{5/3}} &=& \int_0^{\hat\epsilon} \frac{\hat\epsilon^{\prime -\beta}d\hat\epsilon^\prime}{\hat\epsilon^{\prime 1/3}(\hat\epsilon-\hat\epsilon^\prime)^{5/3}} +\int_{\hat \epsilon}^\infty \frac{\hat\epsilon^{\prime -\beta}d\hat\epsilon^\prime}{\hat\epsilon^{\prime 1/3}(\hat\epsilon^\prime-\hat\epsilon)^{5/3}} \nonumber\\
    &=&\hat\epsilon^{-\beta-1}\left(\int_0^1 \frac{(\hat\epsilon^\prime/\hat\epsilon)^{-\beta-1/3}d(\hat\epsilon^\prime/\hat\epsilon)}{(1-\hat\epsilon^\prime/\hat\epsilon)^{5/3}} + \int_1^\infty \frac{(\hat\epsilon^\prime/\hat\epsilon)^{-\beta-1/3}d(\hat\epsilon^\prime/\hat\epsilon)}{(\hat\epsilon^\prime/\hat\epsilon-1)^{5/3}}\right)\nonumber\\
    &=&\hat\epsilon^{-\beta-1}\left(\int_0^1 x^{-\beta-1/3}(1-x)^{-5/3}dx + \int_0^\infty (1+y)^{-\beta-1/3}y^{-5/3} dy\right)\nonumber\\
\end{eqnarray}

\begin{eqnarray}
    I_2=\hat\epsilon^{-\beta}\int_0^\infty \frac{d\hat\epsilon^\prime}{\hat\epsilon^{1/3}|\hat\epsilon-\hat\epsilon^\prime|^{5/3}} &=& \hat\epsilon^{-\beta}\int_0^{\hat\epsilon} \frac{d\hat\epsilon^\prime}{\hat\epsilon^{ 1/3}(\hat\epsilon-\hat\epsilon^\prime)^{5/3}} +\hat\epsilon^{-\beta}\int_{\hat \epsilon}^\infty \frac{d\hat\epsilon^\prime}{\hat\epsilon^{ 1/3}(\hat\epsilon^\prime-\hat\epsilon)^{5/3}} \nonumber\\
    &=& \hat\epsilon^{-\beta-1}\left(\int_0^1 \frac{d(\hat\epsilon^\prime/\hat\epsilon)}{(1-\epsilon^\prime/\hat\epsilon)^{5/3}} + \int_1^\infty \frac{d(\hat\epsilon^\prime/\hat\epsilon)}{(\epsilon^\prime/\hat\epsilon-1)^{5/3}}\right)\nonumber\\
    &=&\hat\epsilon^{-\beta-1}\left(\int_0^1{(1-x)^{-5/3}}dx + \int_0^\infty {y^{-5/3}}{dy}\right)\nonumber\\
\end{eqnarray}
Although the two integrals diverge, they can cancel each other out and result in,
\begin{eqnarray}
    I_1-I_2 &=& \hat\epsilon^{-\beta-1}\left(\int_0^1 x^{-\beta-1/3}(1-x)^{-5/3}dx + \int_0^\infty (1+y)^{-\beta-1/3}y^{-5/3} dy - \int_0^1{(1-x)^{-5/3}}dx - \int_0^\infty {y^{-5/3}}{dy} \right)\nonumber\\
    &=& \hat\epsilon^{-\beta-1}\left(\mathcal{B}(-\beta+\frac{2}{3},-\frac{2}{3}) + \mathcal{B}(-\frac{2}{3}, 1+\beta) -\mathcal{B}(1, -\frac{2}{3})\right)\nonumber\\
    &=&\hat\epsilon^{-\beta-1}\left(\frac{\Gamma(-\beta+\frac{2}{3})\Gamma(-\frac{2}{3})}{\Gamma(-\beta)} + \frac{\Gamma(-\frac{2}{3})\Gamma(1+\beta)}{\Gamma(\frac{1}{3}+\beta)} -\frac{\Gamma(1)\Gamma(-\frac{2}{3})}{\Gamma(\frac{1}{3})}\right)\nonumber\\
     &=& \frac{3\hat\epsilon^{-\beta-1}}{2}\left(1 -  \Gamma(\frac{1}{3})\left(\frac{\Gamma(-\beta+\frac{2}{3})}{\Gamma(-\beta)} + \frac{\Gamma(1+\beta)}{\Gamma(\frac{1}{3}+\beta)} \right) \right)
\end{eqnarray}
One can also rigorously prove that this holds true by changing the integral limit to the corresponding value of $|\Delta|_m$. Then, all the $\mathcal{B}$ functions become incomplete $\mathcal{B}$ functions.
%\end{widetext}

\bibliography{biblio}{}
\bibliographystyle{aasjournal}

\end{document}